\documentclass[%
nofootinbib,
 amsmath,amssymb,
 aps,
 prd,
 twocolumn
]{revtex4-2}

\usepackage{slashed}
\usepackage{graphicx}
\usepackage{dcolumn}
\usepackage{bm}
\usepackage{hyperref}

\usepackage{float}
\usepackage{enumerate}
\usepackage{longtable}
\usepackage{mathrsfs}

\newcommand{\forests}{\mathfrak{F}}
\newcommand{\forestsdeg}{\mathfrak{F}^*}

\newcommand{\nloopsint}{\mathrm{Loop}_0}
\newcommand{\nloopsext}{\mathrm{Loop}_1}
\newcommand{\fermion}{\mathrm{Ferm}}
\newcommand{\coulomb}{\mathrm{Coul}}
\newcommand{\photon}{\mathrm{Ph}}
\newcommand{\edge}{\mathrm{E}}
\newcommand{\ffch}{\mathrm{Ch}}

\newcommand{\ffdeg}{\mathrm{Deg}}
\newcommand{\ffdegz}{\mathrm{Deg}}
\newcommand{\mcmain}{\text{main}}
\newcommand{\mcmin}{\text{min}}
\newcommand{\mcuniform}{\text{uniform}}
\newcommand{\mcmodify}{\text{modify}}
\newcommand{\mcsector}{\text{sector}}

\newcommand{\vsp}{\mathbf}
\newcommand{\vrel}{\mathit}
\newcommand{\schwinger}{\text{Schw}}

\newcommand{\czerosat}{C^0_{\text{sat}}}
\newcommand{\czeromul}{C^0_{\text{mul}}}
\newcommand{\conesat}{C^1_{\text{sat}}}
\newcommand{\cinfsat}{C^{\infty}_{\text{sat}}}
\newcommand{\cinfmul}{C^{\infty}_{\text{mul}}}
\newcommand{\taumin}{\tau_{\text{min}}}
\newcommand{\taumax}{\tau_{\text{max}}}

\begin{document}

\title{
QED vacuum polarization in the Coulomb field of a nucleus: a method of high-order calculation
}

\author{Sergey Volkov}%
 \email{sergey.volkov.1811@gmail.com, volkoff_sergey@mail.ru, svolkov@mpi-hd.mpg.de}
\affiliation{Max~Planck~Institute for Nuclear Physics, Saupfercheckweg~1, D~69117 Heidelberg, Germany
}%


\begin{abstract}
A calculation of the QED vacuum polarization potential in the Coulomb field of a pointlike nucleus was presented in an earlier publication by the author and his collaborators. Corrections up to order $\alpha^2 (Z\alpha)^7$ were evaluated, where $Z$ is the nuclear charge number and $Z\alpha$ is treated as an independent variable. These corrections correspond to two-loop Feynman graphs with proper propagators of fermions in the external field. The calculation employed a reduction to free QED, leading to free QED Feynman graphs with up to eight independent loops. The method of calculation is described here in detail.
\end{abstract}

\maketitle


\section{Introduction}

The energy levels of atomic systems provide stringent tests of bound-state quantum electrodynamics (QED) and its extensions. Each new degree of precision challenges both experiment and theory. It is still often the case that theoretical precision can not reach the experimental bar due to complexity of calculations, especially for high $Z$ (see, for example, ~\cite{indelicato_19}).

Different corrections to the energy levels require distinct methods of calculation. It is reasonable to separate the QED corrections that can be evaluated by treating the nucleus as an infinite-mass source of a classical Coulomb field\footnote{Different corrections of this type may have different factors $f(m,M)$, where $m$ and $M$ are the fermion and nuclear masses.} (see, for example, \cite{eides_grotch_sheluto,yerokhin_15_Hlike}). The QED corrections to this Coulomb potential can be factorized in Feynman graphs in this case. Therefore, they can be evaluated in advance and used in the general calculation after that. 
  

According to a tradition in atomic physics, a potential energy is called a potential (see, for example, ~\cite{blomqvist_1972_vacuum}). In this sense, the Coulomb potential of a pointlike nucleus is given by $V_{\text{Coulomb}}(\vsp{r})=-(Z\alpha)/|\vsp{r}|$. The QED corrections\footnote{Strictly speaking, we use the renormalized $Z\alpha$ when considering QED effects.} to this potential are called the vacuum polarization potential and can be represented ~\cite{blomqvist_1972_vacuum} in the form of a double expansion in $\alpha$ and $Z\alpha$:
$$
V_{\rm VP}(\vsp{r}) = \sum_{{i=1,2,\ldots}\atop
{j = 1,3,\ldots}} V_{ij}(\vsp{r}) \equiv \sum_{{i=1,2,\ldots}\atop
{j = 1,3,\ldots}}
\alpha^i\,(Z\alpha)^j \, \widetilde{V}_{ij}(\vsp{r}).
$$
The summation over $j$ includes only odd $j$'s due to the Furry theorem. Throughout the paper, no interactions with fermions other than electrons and positrons are considered.

Some terms of this expression are known. The leading one-loop term $V_{11}(\vsp{r})$ is the well-known Uehling potential ~\cite{uehling_35}. The remaining part of the one-loop VP potential, $V_{13}(\vsp{r})+V_{15}(\vsp{r})+\ldots$, is the Wichmann-Kroll potential. E.~H.~Wichmann and N.~M.~Kroll studied it and provided ~\cite{wichmann_kroll} useful formulas concerning $\widetilde{V}_{13}(\vsp{r})$. An explicit expression for $\widetilde{V}_{13}(\vsp{r})$ was reported in ~\cite{blomqvist_1972_vacuum}. The Wichmann-Kroll potential can be evaluated in all orders in $Z\alpha$: some procedures suitable for numerical evaluation were presented in ~\cite{soff_88_vp,manakov_89_zhetp} (the case of not pointlike nucleus is also addressed in ~\cite{soff_88_vp}); accurate approximate formulas are also available \cite{fainshtein_91,manakov_12_vgu}.

A semianalytical expression for the leading two-loop term, the K\"all{\'e}n-Sabry potential $V_{21}(\vsp{p})$, was obtained in ~\cite{kaellen_sabry}; here momentum space was used that is defined by
$$
V(\vsp{p})=\int e^{-i\vsp{p}\cdot\vsp{r}}\, V(\vsp{r})\, d^3 \vsp{r}
$$
for any coordinate-space function $V(\vsp{r})$. Explicit approximated coordinate-space formulas for $V_{21}$ are given in ~\cite{blomqvist_1972_vacuum,fullerton_76}. The value of $\widetilde{V}_{23}(\vsp{p}=0)$ was presented in ~\cite{lee_krachkov_2023} in extractable form. 

The author with his collaborators presented the values of $\widetilde{V}_{23}(\vsp{p})$, $\widetilde{V}_{25}(\vsp{p})$, $\widetilde{V}_{27}(\vsp{p})$ for different $|\vsp{p}|$ in tabular form ~\cite{vp_coulomb_potential_2025}. It gave the possibility to resolve one of the most important uncertainties in the theoretical contributions to certain energy levels of high-$Z$ ions ~\cite{vp_coulomb_potential_2025}. See a detailed review concerning potentials and atomic observables in ~\cite{vp_coulomb_potential_2025}. The aim of this paper is to describe the method of calculation of the potentials in detail.

The potentials $V_{2j}$ correspond to two-loop Feynman graphs with proper propagators of fermions in the external Coulomb field. A general method of handling these graphs exists ~\cite{yerokhin_03_epjd,mallampalli_96,goidenko_99_prl,labzowsky_renorm_1996}; it is based on partial-wave decomposition and is nonperturbative in $Z\alpha$ ab initio. That method, for example, allows us to evaluate all 2-loop self-energy Feynman graphs contributing to hydrogenlike ion energy levels. This idea was successfully employed for calculating energy level shifts ~\cite{yerokhin_15_Hlike} as well as the bound-electron $g$-factor ~\cite{sikora_bound_amm_2025}. However, it has not yet been implemented for calculating $\sum_j V_{2j}$ and the corresponding corrections to atomic observables.


In contrast to general-case bound-state problems, the vacuum polarization potential $V_{\rm VP}$ can be treated perturbatively in $Z\alpha$. This means that the methods of free QED can be employed. We use free QED Feynman graphs in our calculation (see Sec. \ref{subsec_feynman_graph}). We apply a method that removes all divergences (infinities) in intermediate values before integration (point by point) in Feynman parametric space (see Sec. \ref{subsec_forest_formula}, \ref{subsec_feynman_parameters}). Dimensional regularization was never used in this calculation as well as other techniques based on expansions of infinities. It allows us to increase the computational efficiency and the accessible number of independent loops in Feynman graphs significantly. On the other hand, it requires a deep understanding of how divergences work in Feynman integrals and how to redistrubute them in order to cancel them according to the rules of physical renormalization. This class of methods has been applied for calculating the free electron anomalous magnetic moment by the author ~\cite{volkov_5loops_total} and other scientists ~\cite{levinewright, carrollyao, kinoshita_6, kinoshita_8_first, kinoshita_10_first, kinoshita_verification_2025}; there exist also applications to other high-order quantum field theory problems ~\cite{kompaniets}.

In contrast to the case of free-electron $g-2$, the free-QED Feynman graphs of $V_{\rm VP}$ do not have infrared (IR) divergences. Therefore, a special subtraction procedure for removing IR divergences is not required: we can apply the BPHZ\footnote{By BPHZ we mean the Bogoliubov-Parasiuk-Hepp-Zimmermann renormalization procedure~\cite{bogolubovparasuk,hepp,zimmerman}.} renormalization almost directly. Applications of BPHZ to the study of two-loop vacuum polarization in the Coulomb field exist~\cite{zschocke_bphz_2002}; however, our procedure yields finite integrals immediately and is suitable for numerical calculation. The only source of IR divergences is the physical QED renormalization conditions; we changed them without affecting the final result to avoid emergent IR divergences (see Sec. \ref{subsec_forest_formula}). One can say that this calculation is a triumph of BPHZ; one should only take care of additional divergent subgraphs that do not exist in the usual free QED (see Sec. \ref{subsec_uv_power_count}, \ref{subsec_forest_formula}). The case $\vsp{p}=0$ also makes difficulties; however, a direct calculation at $\vsp{p}=0$ is not needed for practical purposes (see Sec. \ref{subsec_zero_momentum}). This procedure of working with infinities that was used in the calculation requires a justification; a sketch of this is given in Sec. \ref{subsec_regularization}.

We integrate the obtained Feynman parametric integrals by Monte Carlo. These integrals may have up to 17 independent variables. The integrand landscape can be not so smooth: it may have integrable singularities, acute peaks and so on. We employ a nonadaptive approach similar to the one used in ~\cite{volkov_5loops_total}: the probability density function is constructed using explicit formulas taking into account the combinatorics of the Feynman graph. The presence of two scales (the fermion mass $m$ and external momentum) is an obstacle to obtaining precise results: the value of $|\vsp{p}|/m$ varies from $0.001$ to $100000$. A small region of the integration domain requires a special attention because of that: we use a probability density function dependent on $|\vsp{p}|$ in this small part. See Sec. \ref{sec_integration} for details.

The paper is organized as follows. The method of reduction to finite integrals is explained in Sec. \ref{sec_finite_integrals}. The Monte Carlo integration method and its realization on NVidia GPUs is described in \ref{sec_integration}. The results ~\cite{vp_coulomb_potential_2025} are the product of a very large and complicated computation. They can become reliable only after double-checking. Sec. \ref{sec_individual_results} provides the individual Feynman graph contributions that are necessary for checking.

We use italic style ($\vrel{p}$) for four-vectors, bold face ($\vsp{p}$) for three-vectors. Four-vectors have the form $\vrel{p}=(\vrel{p}_0,\vsp{p}).$ We use the system of units such that $\hbar=c=1$. We use the notation $\slashed{p}=p_{\mu}\gamma^{\mu}$; the tensor $g_{\mu\nu}$ corresponds to the signature $(+,-,-,-)$, and the Dirac matrices fulfill the condition $\gamma_{\mu}\gamma_{\nu}+\gamma_{\nu}\gamma_{\mu}=2g_{\mu\nu}$.

\section{Reduction to finite integrals}\label{sec_finite_integrals}

\subsection{Feynman graphs}\label{subsec_feynman_graph}

\begin{figure}
\includegraphics[width=85mm]{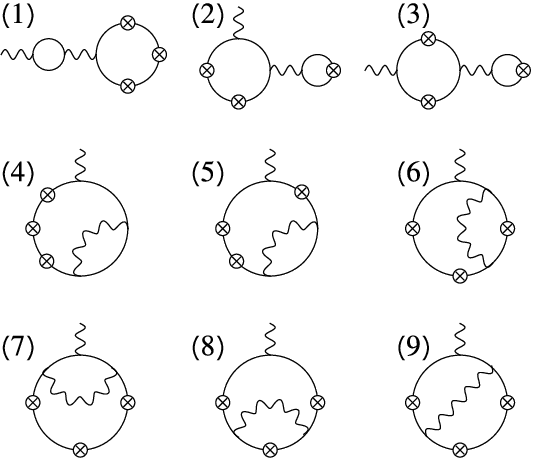}
\caption{All Feynman graphs contributing to $V_{23}$.
Solid lines denote free-fermion propagators,
wavy lines denote photon propagators, and circled crosses
denote Coulomb interactions with the nucleus.
\label{fig23}
}
\end{figure}

\begin{figure}
\includegraphics[width=85mm]{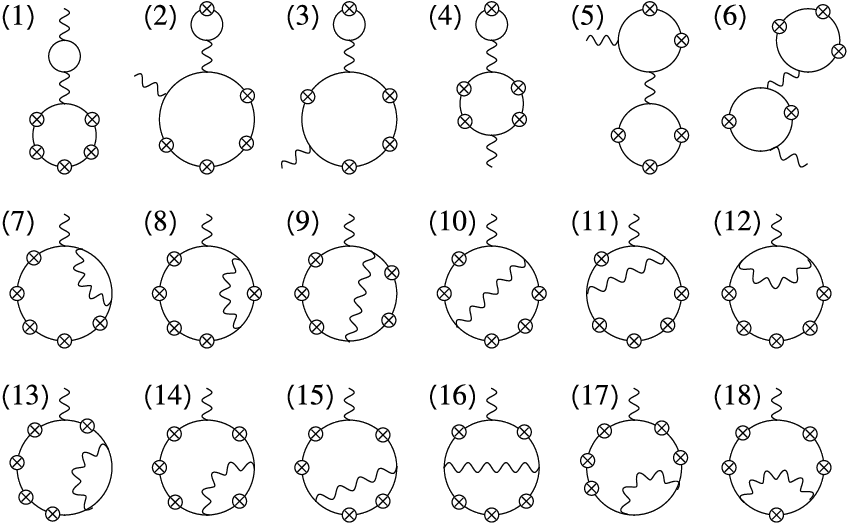}
\caption{All Feynman graphs contributing to $V_{25}$.
\label{fig25}
}
\end{figure}

\begin{figure}
\includegraphics[width=85mm]{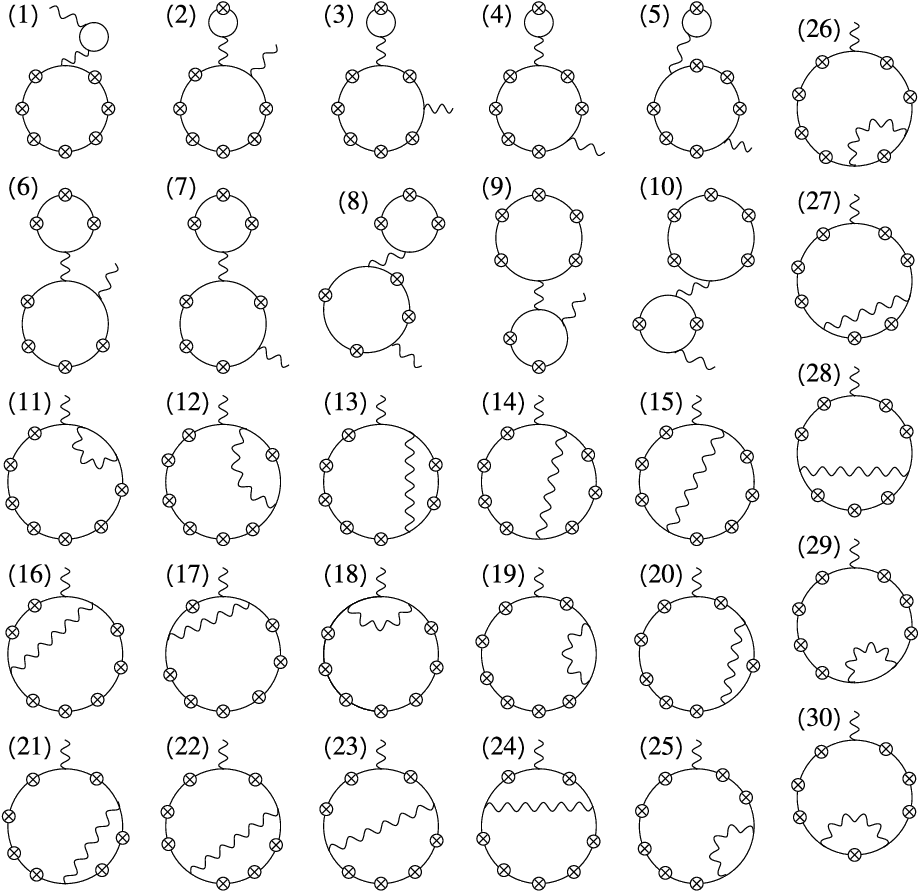}
\caption{All Feynman graphs contributing to $V_{27}$.
\label{fig27}
}
\end{figure}

The Feynman graphs contributing to $\widetilde{V}_{23}$, $\widetilde{V}_{25}$, and $\widetilde{V}_{27}$ are shown in Fig.~\ref{fig23}, Fig.~\ref{fig25}, and Fig.~\ref{fig27}, respectively. The contributions of the graphs labeled (1) in all figures can be expressed as products of two one-loop values in momentum space. The remaining graphs are considered irreducible two-loop graphs. 

\begin{figure}
\includegraphics[width=85mm]{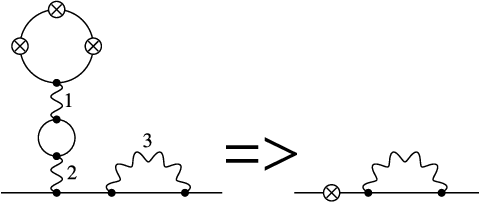}
\caption{An example of correcting the Coulomb potential in a Feynman graph. Left side: a Feynman graph containing (1) in Fig. \ref{fig23} as a subgraph. Right side: the equivalent graph containing an interaction with a corrected potential.
\label{fig_insert_example}
}
\end{figure}

An example of correcting the Coulomb potential is shown in Fig. \ref{fig_insert_example}. The left side contains a Feynman graph with Coulomb interactions and (1) from Fig. \ref{fig23} as a subgraph, whereas the right side is the equivalent graph with corrected interactions. The renormalization is treated in a way that makes this replacement consistent. In general, photons 1, 2, 3 in Fig. \ref{fig_insert_example} (left) can be treated as internal photons in the same way (for instance, using the Feynman gauge or any other gauge). The graphs with corrected interactions have a restriction: they do not contain an internal photon that separates some subset of the corrected-interaction vertices from the external lines of this graph. Since the paper is restricted to the two-loop case, we will not consider this combinatorics here in detail.

\begin{figure}
\includegraphics[width=85mm]{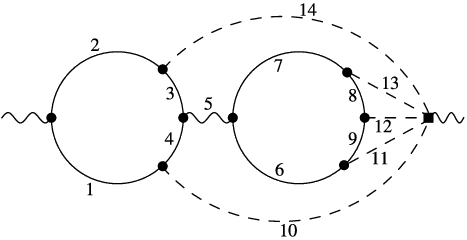}
\caption{The unfolded Feynman graph corresponding to graph (6) in Fig. \ref{fig25}.}
\label{figdisclose}
\end{figure}

We \emph{unfold} each Feynman graph in order to apply free-QED methods. Unfolding means replacing each Coulomb interaction with a propagator line attached to the same additional vertex. This new vertex has a fictitious external line that carries the same momentum as the original external line.
An example of unfolding is shown in Fig. \ref{figdisclose}. The added vertex, marked by a boxed dot, embodies momentum conservation for Coulomb lines; it does not contribute any factor and carries no tensor index. We refer to it as the \emph{fictitious vertex}. Dashed lines denote Coulomb propagators. The Feynman gauge is used; the photon, fermion, and Coulomb propagators are given by
\begin{equation}\label{eq_propagators}
\frac{i(\slashed{\vrel{q}}+m)}{\vrel{q}^2-m^2+i0},\quad \frac{-ig_{\mu\nu}}{\vrel{q}^2+i0},\quad \frac{-\delta(\vrel{q}_0)\delta_{\mu 0}}{\vsp{q}^2},
\end{equation}
respectively (up to coefficient), where $m$ is the fermion mass. The unfolded graphs for $\widetilde{V}_{23}$, $\widetilde{V}_{25}$, and $\widetilde{V}_{27}$ contain 4, 6, and 8 independent loops, respectively. We use wavy lines for both external lines (main and fictitious) as well as for the photon internal lines.

The goal of the unfolding is to obtain a convenient representation with unified internal and external momenta integrations. We will use it for an analysis of the UV behavior.

The unfolded-graph Feynman amplitude $F_{\mu}(p)$ is defined for all $p$ satisfying $p_0=0$ and is a number. After summation over all graphs of the required order, it satisfies $p^{\mu}F_{\mu}(p)=0$. This can be shown by a usual diagrammatic procedure used in QED for proving the Ward-Takahashi identities\footnote{See, for example, ~\cite{peskin}, Section 7.4 ``The Ward-Takahashi Identity''.}. Combined with symmetry properties, this implies that $F_{\mu}(\vsp{p})=0$ for $\mu\neq 0$. Therefore, $V_{ij}(\vsp{p})$ is extracted from $F_0(\vsp{p})$. The photon propagators in (\ref{eq_propagators}) correspond to the Feynman gauge. Using the same diagrammatic methods, one can show that $F_{\mu}(\vsp{p})$ is independent of this gauge choice. The Coulomb-propagator gauge can likewise be chosen arbitrarily, by the same argument. This gauge choice is independent of the photon-propagator gauge.

\subsection{Ultraviolet power counting}\label{subsec_uv_power_count}

We now perform a preliminary analysis of UV divergences in unfolded Feynman graphs. Let $s$ be an arbitrary set of internal lines. Suppose the momenta of the lines in $s$ are variable, while the others are fixed. By $\coulomb(s)$ and $\fermion(s)$ we denote the sets of all Coulomb and fermion lines in $s$, respectively. By $\nloopsint(s)$ and $\nloopsext(s)$ we denote the number of independent loops in $s\backslash\coulomb(s)$ and $s$, respectively. The momentum-space integration domain can be parametrized by $\nloopsint(s)$ four-dimensional variables and $\nloopsext(s)-\nloopsint(s)$ three-dimensional variables, for a total of
\begin{gather*}
4\times\nloopsint(s) + 3\times(\nloopsext(s)-\nloopsint(s)) \\
= 3\times\nloopsext(s)+\nloopsint(s)
\end{gather*}
dimensions. 
The examined integral is (superficially) UV divergent if $\omega(s)\geq 0$, where
\begin{equation}\label{eq_uv_div_index}
\omega(s) = \frac{1}{2} \nloopsint(s) + \frac{3}{2} \nloopsext(s) + \frac{1}{2}|\fermion(s)| - |s|.
\end{equation}
For example, for Fig. \ref{figdisclose} we find
\begin{gather*}
\omega(\{1,2,3,4\})=0,\ \ \omega(\{5,8\})=-1.5, \\
\omega(\{8,9,11,12,13\})=-1,\ \ \omega(\{6,7,8,9,11,12\})=-0.5.
\end{gather*}
Let us analyze this expression further. Two lines $a,b\in s$ are said to \emph{lie in the same component of $s$} if there exists a path connecting an endpoint of $a$ with an endpoint of $b$ that does not pass through the fictitious vertex. This relation splits $s$ into components. For example, in Fig. \ref{figdisclose}, the set $\{8,11,12\}$ splits into two components: $\{11\}$ and $\{8,12\}$, whereas the set $\{3,4,7,9,10,14\}$ splits into three: $\{7\}$, $\{9\}$, $\{3,4,10,14\}$.

Suppose $s$ has only one component. Define
\begin{gather*}
E=|s|,\quad F=|\fermion(s)|,\quad Q=|\coulomb(s)|, \\
 L_0=\nloopsint(s),\quad L_1=\nloopsext(s).
\end{gather*}
Let $V$ be the number of vertices incident\footnote{A vertex $v$ is \emph{incident} to a line $l$, if $v$ is one of the endpoints of $l$.} to $s$ (excluding the fictitious vertex). Let $N_Q$ denote the number of Coulomb lines external to $s$ (attached to a nonfictitious vertex incident to a line of $s$). Similarly, $N_{\gamma}$ and $N_f$ denote the numbers of external photon lines (including the external line of the full graph, but not the fictitious external line) and external fermion lines relative to $s$, respectively; a line may be counted twice, if both its endpoints are incident to lines in $s$. Finally, let $I=1$ if the fictitious vertex is incident to some line in $s$, and $I=0$ otherwise. For example, for $s=\{6,7,8,9,11,12\}$ in Fig. \ref{figdisclose} we have
\begin{gather*}
E=6,\ F=4,\ Q=2,\ L_0=1,\ L_1=2,\\
N_Q=1,\ N_{\gamma}=1,\ N_f=0,\ I=1,
\end{gather*}
while for $s=\{2,3,4,5\}$ we have
\begin{gather*}
E=4,\ F=3,\ Q=0,\ L_0=0,\ L_1=0,\\ 
N_Q=2,\ N_{\gamma}=1,\ N_f=4,\ I=0
\end{gather*}
(the fermion line 1 was counted twice in $N_f$).
Examining the ultraviolet degree of divergence of $s\backslash\coulomb(s)$ in the usual QED way, we obtain
$$
2L_0 + \frac{1}{2}F-(E-Q) = 2 - \frac{1}{2}(N_{\gamma}+N_Q+Q) - \frac{3}{4}N_F.
$$
Taking into account that $L_1-L_0=Q-I$, we find
$$
\omega(s)=2-\frac{3}{2}I - \frac{1}{2}(N_Q+N_{\gamma}) - \frac{3}{4}N_F.
$$
If $s$ consists of several components, these values are summed over components. The case $I=0$ corresponds to standard QED. In this case there are four types of subgraphs with $\omega\geq 0$: \emph{fermion self-energy} ($N_f=2,\ N_Q=N_{\gamma}=0$); \emph{vertexlike} ($N_f=2,\ N_Q+N_{\gamma}=1$); \emph{photon self-energy} ($N_f=0,\ N_Q+N_{\gamma}=2$); \emph{photon-photon scattering} ($N_f=0,\ N_Q+N_{\gamma}=4$).

Now consider the case $I=1$ with $\omega(s)\geq 0$. The situation $N_{\gamma}=0$ is excluded, since the original graph was connected prior to unfolding. The only admissible values are
$$
N_Q=0,\quad N_{\gamma}=1,\quad N_f=0.
$$
We call such one-component sets \emph{potential subgraphs}. All of them must be treated to eliminate UV divergences. The full graph is always potential. In Fig. \ref{figdisclose}, for example, the set $\{6,7,8,9,11,12,13\}$ is a potential subgraph, in addition to the full graph.

\subsection{Divergence elimination and renormalization via a forest formula}\label{subsec_forest_formula}

We work with an unfolded graph and its subgraphs, considering only those that are one-particle irreducible and contain all lines connecting the vertices of the given subgraph.

Two subgraphs are said to \emph{overlap} if neither is contained in the other and their sets of internal lines sets have a nonempty intersection.

A set of subgraphs of a graph is called a \emph{forest} if no two of its elements overlap.

For an unfolded graph $G$, we denote by $\forests[G]$ the set of all forests $F$ consisting of self-energy, vertexlike, photon-photon scattering, or potential subgraphs of $G$ and satisfying the condition $G\notin F$. For example, for $G$ from Fig. \ref{figdisclose}, the set $\forests[G]$ is exactly the set of all subsets of $\{\langle 1,2,3,4\rangle,\ \langle 6,7,8,9 \rangle,\ \langle 6,7,8,9,11,12,13 \rangle\}$ (8 subsets); here subgraphs are denoted by the enumeration of internal lines.

To obtain a divergence-free, properly renormalized Feynman amplitude for each unfolded graph $G$, we apply Zimmermann's forest formula ~\cite{zavialovstepanov,zimmerman}
\begin{equation}\label{eq_forest_formula}
\sum_{F=\{G_1,\ldots,G_n\}\in\forests[G]} (-1)^n S_{G_1}S_{G_2}\ldots S_{G_n}
\end{equation}
to it.
Each term in this expression should be interpreted as a sequential transformation of the subgraph $G_j$ Feynman amplitudes with the linear operators $S_{G_j}$. The subgraphs are ordered from smaller to larger (with respect to inclusion), and the transformed amplitudes of smaller subgraphs are taken into account in the transformations of the larger ones. 

The linear operators $S_{G'}$ are defined differently for different types of subgraphs $G'$:
\begin{enumerate}
\item If $G'$ is potential, its Feynman amplitude is $\Pi_{\mu}(p)$, where $p$ is its external momentum entering the nonfictitious vertex, and $\mu$ is the corresponding tensor index. In this case, we put
$$
[S\Pi]_{\mu}(p)=\Pi_{\mu}(0).
$$
\item If $G'$ is a photon-photon scattering subgraph, its Feynman amplitude is $\Pi_{\mu_1 \mu_2 \mu_3 \mu_4}(p_1,p_2,p_3,p_4)$. Analogously, we put
$$
[S\Pi]_{\mu_1 \mu_2 \mu_3 \mu_4}(p_1,p_2,p_3,p_4)=\Pi_{\mu_1 \mu_2 \mu_3 \mu_4}(0,0,0,0).
$$
\item If $G'$ is a photon self-energy subgraph, its Feynman amplitude is $\Pi_{\mu\nu}(p)$. We can take the quadratic Taylor expansion
$$
[S\Pi]_{\mu\nu}=\Pi_{\mu\nu}(0)+\frac{1}{2} \left.\frac{\partial^2 \Pi_{\mu\nu}(p)}{\partial p_{\xi}\,\partial p_{\eta}}\right|_{p=0} p_{\xi}p_{\eta}.
$$
However, it is more efficient to utilize the substitution
$$
\Pi_{\mu\nu}(p) \rightarrow p^2 g_{\mu\nu}[h_2(0)-h_2(p^2)]
$$
instead of $(1-S)\Pi$ in the forest formula, where $\Pi_{\mu\nu}(p)=h_1(p^2)g_{\mu\nu}+h_2(p^2) p_{\mu} p_{\nu}$. This trick, used in our calculation, shortens the expressions and is common in the literature; see, for example, ~\cite{kinoshita_infrared}.
\item If $G'$ is vertexlike, its Feynman amplitude is $\Gamma_{\mu}(p,q)$, where the incoming and outgoing fermion momenta are $p-\frac{q}{2}$ and $p+\frac{q}{2}$, and the photon momentum is $q$. Its $q=0$ part can be written as
\begin{gather*}
\Gamma_{\mu}(p,0) \\
=a(p^2)\gamma_{\mu}+b(p^2)p_{\mu}+c(p^2)\slashed{p}p_{\mu} + d(p^2) (\slashed{p}\gamma_{\mu}-\gamma_{\mu}\slashed{p}).
\end{gather*}
In this case, we put
\begin{equation}\label{eq_operator_vertex}
[S\Gamma]_{\mu}(p,q)=a((m_0)^2)\gamma_{\mu},
\end{equation}
where $(m_0)^2$ is an arbitrary constant. In our calculation ~\cite{vp_coulomb_potential_2025}, we take $(m_0)^2=-5m^2$ to minimize cancellations inside integrals, but the individual graph contributions for different values of $(m_0)^2$ are given in Sec. \ref{sec_individual_results}.
\item If $G'$ is a fermion self-energy subgraph, its Feynman amplitude is 
$$
\Sigma(p)=u(p^2)+v(p^2)\slashed{p}.
$$
In this case, we put
\begin{equation}\label{eq_operator_fse}
[S\Sigma](p)=u(m^2)+v(m^2)m+v((m_0)^2)(\slashed{p}-m),
\end{equation}
where $(m_0)^2$ is the same as for vertexlike subgraphs.
\end{enumerate}

It is well known that the photon-photon scattering amplitude $\Pi$ satisfies
$$
\Pi_{\mu_1 \mu_2 \mu_3 \mu_4}(0,0,0,0)=0
$$
after summation over all graphs, due to QED gauge invariance. Thus, these subtractions for photon-photon scattering subgraphs are fictitious, but they are necessary to remove UV divergences in individual graphs. The same reasoning applies to potential subgraphs.

The operator $S$ for fermion self-energy and vertexlike subgraphs preserves the Ward identity and extracts the mass part completely; unlike the standard on-shell renormalization, it does not produce IR divergences; see ~\cite{volkov_2015,volkov_method_details_2023}. In general, the difference introduces a wave function renormalization factor in the final result\footnote{This statement is an important component of the QED renormalizability proof. See, for example, ~\cite{collins}, its Chapter 5 ``Renormalization'', Section 5.6 ``Relation to $\mathscr{L}$''. A clear explanation is also given in ~\cite{bogoliubov_shirkov}. See also an application example of changing the subtraction point in the Section 19.9 ``Analytic Continuation and Intermediate Renormalization'' of ~\cite{bjorkendrell2}.}. However, in our situation, no such factor appears because there are no external fermions. 

To obtain the final result, one must sum the contributions of all graphs from Fig. \ref{fig23}, \ref{fig25}, or \ref{fig27} (except graph 1), renormalized in this way, and add the contribution of graph 1 obtained from lower-order values. No residual renormalization is required.

Zimmermann's forest formuls itself does not explain how to obtain a finite integral for each graph; this is explained in terms of Feynman parameters in Sec. \ref{subsec_feynman_parameters}. Note that in this version of Zimmermann's forest formula, the full graph $G$ is not subtracted; this subtraction is implemented explicitly in Sec. \ref{subsec_feynman_parameters} for a proper definition in Feynman parametric space.

\subsection{Constructing Feynman-parametric integrals}\label{subsec_feynman_parameters}

Suppose the unfolded graph $G$ has $n$ internal lines, enumerated by $1,2,\ldots,n$.

The procedure for obtaining a Feynman-parametric integral for an unfolded graph can be explained conveniently using Schwinger parameters as an intermediate representation:
\begin{itemize}
\item For arbitrary values of the Schwinger parameters $z_1,\ldots,z_n\geq 0$ and an infrared regulator $\varepsilon>0$, introduce the Schwinger parametric propagators
$$
(\slashed{\vrel{q}}+m)e^{iz_j(\vrel{q}^2-m^2+i\varepsilon)},\quad ig_{\mu\nu}e^{iz_j (\vrel{q}^2+i\varepsilon)},
$$
$$
i\delta(\vrel{q}_0)\delta_{\mu 0} e^{iz_j(\vsp{q}^2+i\varepsilon)}
$$
instead of (\ref{eq_propagators}), where $j$ labels the internal line number.
\item Obtain $F^{\schwinger}(\vsp{p},z_1,\ldots,z_n,\varepsilon)$ as the result of integrating over the loop momenta with these propagators. We assume that the forest formula (\ref{eq_forest_formula}) is applied. It can be derived using known explicit formulas for integrals of multidimensional Gaussian functions multiplied by polynomials (ignoring the divergence); see, for example, ~\cite{bogoliubov_shirkov,zavialov,smirnov}. 
We obtain
\begin{gather*}
F^{\schwinger}(\vsp{p},z_1,\ldots,z_n,\varepsilon)=\frac{1}{D_0(z)^{1/2}D_1(z)^{3/2}} \\
\times \left( \sum_{l=0}^{L/2} R_l(\vsp{p}^2,z) \right) e^{iA(z)\vsp{p}^2+iB(z)-\varepsilon \sum_{j=1}^n z_j} \\
+\text{counterterms},
\end{gather*}
where $D_0$ and $D_1$ are the first Symanzik polynomials of the original and unfolded graph respectively; $R_l$ is a polynomial in $\vsp{p}^2$ (constant for $l=L/2$), rational and homogeneous of degree $-l$ in $z$; $L$ is the number of fermion lines in the graph; $A$ and $B$ are rational, homogeneous of degree $1$, and real; the counterterms correspond to the case $F\neq\emptyset$ in (\ref{eq_forest_formula}) and have the same form but with different functions $D_j,R_j,A$.
\item Pass to Feynman parameters via
$$
F(\vsp{p},z)=\frac{1}{\vsp{p}^2}\lim_{\varepsilon\rightarrow +0}
$$
$$\int_0^{+\infty} \lambda^{n-1} \left( F^{\schwinger}(\vsp{p},\lambda z,\varepsilon)-F^{\schwinger}(0,\lambda z,\varepsilon)\right) d\lambda.
$$
Here, division by $\vsp{p}^2$ accounts for the propagator of the external photon (which must be included in the potential as an internal photon propagator); the subtraction implements the remaining subtraction of the full graph in Zimmermann's forest formula.
\end{itemize}
This integral and the limit can be evaluated analytically with the help of the following formulas:
\begin{gather*}
\int_{0}^{+\infty} e^{a\lambda} \lambda^n d\lambda = \frac{n!}{(-a)^{n+1}}, \\
\int_{0}^{+\infty} (e^{a\lambda}-e^{b\lambda}) \lambda^{-1} d\lambda = \log(-b)-\log(-a).
\end{gather*}
We obtain
\begin{equation}\label{eq_integrand}
\begin{array}{c}
F(\vsp{p},z) = \frac{1}{D_0(z)^{1/2}D_1(z)^{3/2}} \\
\times\left[ \frac{W_0(z)}{\vsp{p}^2} \log\left(1+\frac{A(z)}{B(z)}\vsp{p}^2\right) \right. \\
+\sum_{l=1}^{L/2} \left( \frac{W_l(z)}{\vsp{p}^2} \left( \frac{1}{(A(z)\vsp{p}^2+B(z))^l}-\frac{1}{B(z)^l} \right)\right) \\
\left. 
+\sum_{l=1}^{L/2} \frac{Y_l(\vsp{p}^2,z)}{(A(z)\vsp{p}^2+B(z))^l}\right]+\text{counterterms},
\end{array}
\end{equation}
where $W_l(z)$ are rational in $z$; $Y_l(\vsp{p}^2,z)$ are polynomial in $\vsp{p}^2$ and rational in $z$. Each $W_l(z)$ can be easily extracted from the constant term of $R_{\frac{L}{2}-l}(\vsp{p}^2,z)$ (in $\vsp{p}^2$); each $Y_l(\vsp{p}^2,z)$ comes from the remaining terms of $R_{\frac{L}{2}-l}(\vsp{p}^2,z)$. The $\log$-term arises due to a logarithmic overall UV divergence of the full Feynman graph; the full graph subtraction separation ensures that these $\log$-terms are finite.

The contribution of the graph can then be evaluated by integrating
\begin{equation}\label{eq_feynman_parametric}
\int_{z>0} F(\vsp{p},z_1,\ldots,z_n)\,\delta(z_1+\ldots+z_n-1)\,d^n z
\end{equation}
numerically.

\subsection{Divergence at $\vsp{p}=0$}\label{subsec_zero_momentum}

We can define $F(\vsp{p}=0,z)$ by
\begin{gather*}
F(\vsp{p}=0,z)=\lim_{\vsp{p}\rightarrow 0} F(\vsp{p},z) \\
= \frac{1}{D_0(z)^{1/2}D_1(z)^{3/2}} \times \left( \frac{A(z)}{B(z)} \times \left[ W_0(z) - \sum_{l=1}^{L/2} \frac{lW_l(z)}{B(z)^{l}} \right]\right. \\
\left. +\sum_{l=1}^{L/2} \frac{Y_l(0,z)}{B(z)^l} \right)+\text{counterterms}. 
\end{gather*}

The factor $A(z)/B(z)$ renders the integral (\ref{eq_feynman_parametric}) divergent even for Wichmann-Kroll contributions $V_{13}$ and $V_{15}$. Let us illustrate this with examples.

One way to test the integral (\ref{eq_feynman_parametric}) for divergence is to consider all possible nontrivial subsets $s$ of internal lines, set $z_j=\delta a_j$ for $j\in s$ and $z_j=a_j$ for $j\notin s$, with $a_j>0$ fixed, and analyze the asymptotic behavior of $F(\vsp{p},z)$ as $\delta\rightarrow 0$. If $|F(\vsp{p},z)\delta^{|s|}|=\Omega(1)$ as $\delta\rightarrow 0$, then the integral diverges.

\begin{figure*}
\includegraphics[width=85mm]{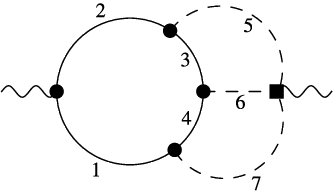}
\ \ \ \ \ \ 
\includegraphics[width=70mm]{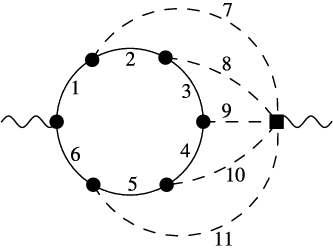}
\caption{Unfolded Feynman graphs of $\widetilde{V}_{13}$ (left) and $\widetilde{V}_{15}$ (right)
\label{fig_v13_v15}
}
\end{figure*}

Consider the unfolded graph of $V_{13}$ in Fig. \ref{fig_v13_v15} (left) with $s=\{1,2,3,4\}$. Expression (\ref{eq_integrand}) contains a main term and a counterterm related to the subtraction of a photon-photon scattering subgraph. For $\vsp{p}\neq 0$, both terms contribute at asymptotic order $1$ to $F(\vsp{p},z)\delta^{|s|}$ as $\delta\rightarrow 0$ (since $s$ is logarithmically divergent). However, the contributions cancel, leaving order $\delta$. Because $s$ contains all fermion lines, $B(z)$ is of asymptotic order $\delta$. To analyze $A(z)$, one can use the analogy with electric circuits\footnote{See ~\cite{bjorkendrell2}, Chapter 18 ``Dispersion Relations,'' Section 18.4 ``Generalization to Arbitrary Graphs and the Electrical Circuit Analogy.''}: $A(z)$ is the resistance of the graph with line resistances $z$; since $s$ does not contain a path between the external lines, $A(z)$ is of asymptotic order $1$, both for the main term and for the counterterm. Thus, the factor $A(z)/B(z)$ of order $\delta^{-1}$ sets the asymptotic order of $|F(\vsp{p}=0,z)\delta^{|s|}|$ to $1$, making the corresponding integral (\ref{eq_feynman_parametric}) divergent.

This divergence at $\vsp{p}=0$ is not always associated with a UV-divergent subgraph. For example, in $V_{15}$ (Fig. \ref{fig_v13_v15}, right) with $s=\{1,2,3,4,5,6\}$, the factor $A(z)/B(z)$, again of order $\delta^{-1}$, changes the asymptotic behavior of $|F(\vsp{p},z)\delta^{|s|}|$ from order $\delta$ to order $1$ at $\vsp{p}=0$.

From the momentum-space point of view, the divergence comes from an integration with respect to Coulomb momenta in the IR limit.

This divergence disappears after summing over Feynman graphs. This can be showed by a more careful consideration of 4-photon and 6-photon Ward identities, but the details are beyond the scope of the present paper. We do not calculate the potential exactly at $\vsp{p}=0$. Instead we extrapolate from values near $0$. On the other hand, for individual graphs the divergence is real, and some contributions tend to $\infty$ as $\vsp{p}\rightarrow 0$; see Sec. \ref{sec_individual_results}.

\subsection{Correctness of the regularization}\label{subsec_regularization}

The final value of $\widetilde{V}_{ij}(\vsp{p})$ is obtained as
\begin{equation}\label{eq_summation}
\begin{array}{c}
\widetilde{V}_{ij}(\vsp{p}) = \sum_{r} \\ \int_{z>0} F_r(\vsp{p},z_1,\ldots,z_n) \delta(z_1+\ldots+z_n-1) d^n z+ \\ \text{products of lower-order terms},
\end{array}
\end{equation}
where the summation is taken over the Feynman graphs of $\widetilde{V}_{ij}$ (excluding products of lower-order terms), and $F_r$ is defined by (\ref{eq_integrand}) for the $r$-th graph. Since we use a special procedure to remove divergences, its correctness must be justified. 

There is no formal definition of correct regularization in quantum field theory. However, dimensional regularization can be used as a specimen, since it has repeatedly been shown to be equivalent to other formulations, produce reliable results and preserve essential properties. Suppose we work in $D=(3-\varepsilon)+1$ dimensions. We define the dimensionally regularized integrand for one graph analogously to the $D=4$ case:
\begin{equation}\label{eq_integrand_dimreg}
\begin{array}{c}
F(\vsp{p},z,\varepsilon) = \frac{1}{D_0(z)^{1/2}D_1(z)^{(3-\varepsilon)/2}} \\
\times\left[ \sum_{l=0}^{L/2} \left( \frac{W_l(z,\varepsilon)}{\vsp{p}^2} \left( \frac{1}{(A(z)\vsp{p}^2+B(z))^{l+k\varepsilon}}-\frac{1}{B(z)^{l+k\varepsilon}} \right)\right) \right. \\
\left. +\sum_{l=1}^{L/2} \frac{Y_l(\vsp{p}^2,z,\varepsilon)}{(A(z)\vsp{p}^2+B(z))^{l+k\varepsilon}}\right]+\text{counterterms};
\end{array}
\end{equation}
here $k=L_1/2$, where $L_1$ is the number of independent loops in the unfolded graph; the functions $W_l(z,\varepsilon)$ and $Y_l(\vsp{p}^2,z,\varepsilon)$ satisfy the same properties with respect to $\vsp{p}^2$ and $z$ as in the $D=4$ case. It is straightforward to check that
$$
\lim_{\varepsilon\rightarrow 0} W_l(z,\varepsilon) = W_l(z),\quad \lim_{\varepsilon\rightarrow 0} Y_l(\vsp{p}^2,z,\varepsilon) = Y_l(\vsp{p}^2,z)
$$
for $l\geq 1$, and
$$
-k\lim_{\varepsilon\rightarrow 0} \varepsilon W_0(z,\varepsilon) = W_0(z).
$$

Now fix $\vsp{p}\neq 0$. The integral (\ref{eq_feynman_parametric}), applied to $F(\vsp{p},z,\varepsilon)$, exists for all $\varepsilon$ such that $|\varepsilon|<\varepsilon_0$ for some $\varepsilon_0>0$. Moreover,
\begin{gather*}
\lim_{\varepsilon\rightarrow 0} \int_{z>0} F(\vsp{p},z_1,\ldots,z_n,\varepsilon) \delta(z_1+\ldots+z_n-1) d^n z \\ 
= \int_{z>0} F(\vsp{p},z_1,\ldots,z_n) \delta(z_1+\ldots+z_n-1) d^n z.
\end{gather*}

By definition, set
\begin{gather*}
\widetilde{V}^{\text{DR,1}}_{ij}(\vsp{p},\varepsilon) = \sum_{r} \\
\int_{z>0} F_r(\vsp{p},z_1,\ldots,z_n,\varepsilon) \delta(z_1+\ldots+z_n-1) d^n z \\
+\text{products of lower-order terms},
\end{gather*}
where $F_r$ is defined by (\ref{eq_integrand_dimreg}) for the $r$-th Feynman graph. 

Let $\widetilde{V}^{\text{DR,2}}_{ij}(\vsp{p},\varepsilon)$ denote the value  defined without counterterm subtractions in the integrands, but with explicit perturbative renormalization of the constants. This definition may require an additional auxiliary regularization and analytic continuation of intermediate functions (we will not consider it in detail, since formulations of this kind are commonly used in calculations). We will use $\widetilde{V}^{\text{DR,2}}_{ij}$ as a specimen. One can prove that
$$
\widetilde{V}^{\text{DR,1}}_{ij}(\vsp{p},\varepsilon)=\widetilde{V}^{\text{DR,2}}_{ij}(\vsp{p},\varepsilon)
$$ 
for all $\varepsilon$ with $0<|\varepsilon|<\varepsilon_0$. To do this, all power of dimensional regularization can be used: the ability to change the order of momentum integrations, perform a linear change of integration variables, the preservation of Ward identities, independence of auxiliary regularization and so forth. As a consequence,
$$
\lim_{\varepsilon\rightarrow 0}\widetilde{V}^{\text{DR,2}}_{ij}(\vsp{p},\varepsilon) = \widetilde{V}_{ij}(\vsp{p}),
$$
where $\widetilde{V}_{ij}(\vsp{p})$ is defined by (\ref{eq_summation}).

\section{Integration}\label{sec_integration}

\subsection{General idea}

To integrate (\ref{eq_feynman_parametric}), we use predefined probability density functions of the form
$$
g(z,\vsp{p})=C_{\mcmain}\times g_0(z,\vsp{p})+\text{stability terms},
$$
where $0\leq C_{\mcmain}\leq 1$ is an arbitrary number, and the function $g_0$ has the form
$$
g_0(z,\vsp{p}) = C(\vsp{p}) \times \frac{\prod_{l=2}^n \left( z_{j_l} / z_{j_{l-1}} \right)^{\ffdegz(\{j_l,j_{l+1},\ldots,j_n\},\vsp{p})}}{z_1 z_2 \ldots z_n},
$$
where
\begin{itemize}
\item $(j_1,\ldots,j_n)$ is a permutation of $1,\ldots,n$ such that $z_{j_1}\geq z_{j_2}\geq\ldots\geq z_{j_n}$ (this partition of the integration domain is known as the Hepp sectors ~\cite{hepp});
\item $\ffdegz(s,\vsp{p})$ are arbitrary positive real numbers defined for any set $s\subseteq \{1,2,\ldots,n\}$ (excluding the empty and full sets);
\item $C(\vsp{p})$ is chosen such that
\begin{equation}\label{eq_dens_normalization}
\int_{z_1,\ldots,z_n>0} g_0(z,\vsp{p}) \delta(z_1+\ldots+z_n-1) dz_1\ldots dz_n = 1.
\end{equation}
\end{itemize}
The stability terms are described in Sec. \ref{subsec_stability}. The idea of these approximations in Hepp's sectors originates from ~\cite{speer}. Probability density functions of this type were used by the author in the free electron $g-2$ calculation ~\cite{volkov_5loops_total}. A method for fast random sample generation was described in  ~\cite{volkov_prd}; a similar algorithm was also implemented in \texttt{feyntrop} ~\cite{borinsky_quadrature,borinsky_minkowsky}.

\subsection{Obtaining $\ffdegz(s,\vsp{p})$ for each set $s$ and momentum $\vsp{p}$}\label{subsec_degz}

Let $G$ be an unfolded Feynman graph of $\widetilde{V}_{2j}$. Denote by $\edge[G]$  the set of all internal lines of $G$; we label them $1,2,\ldots,n$. We consider all sets $s\subseteq\edge[G]$. We will use the symbol $G$ implicitly in the definitions of this section.

For a forest $F$, a graph $G''\in F$ is called a \emph{child} of $G'\in F$ in $F$ if $G''$ is a maximal (with respect to inclusion) element of $F$ properly contained in $G'$.
For example, if $G$ is the graph in Fig. \ref{figdisclose}, 
\begin{gather*}
G_1=\langle 1,2,3,4\rangle,\quad G_2=\langle 6,7,8,9\rangle,\\
G_3=\langle 6,7,8,9,11,12,13\rangle
\end{gather*}
(subgraphs are given by enumeration of the internal lines), then $G_1$ and $G_3$ are the only children of $G$ in $\{G,G_1,G_2,G_3\}$, and $G_2$ is a child of $G_3$; however, $G_1$ and $G_2$ are children of $G$ in $\{G,G_1,G_2\}$.

\begin{figure*}
\includegraphics[width=130mm]{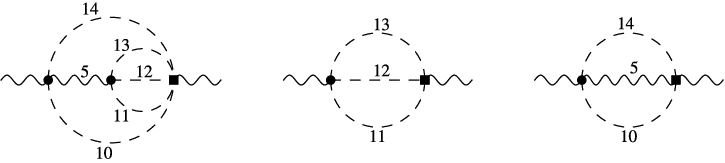}
\caption{Graphs obtained from Fig. \ref{figdisclose} by some operations.
\label{fig_shrink}
}
\end{figure*}

If $F\in\forests[G]$ and $G'\in F$, then $G'/F$ denotes the graph obtained from $G'$ by shrinking all children of $G'$ in $F$ to points. For example, for $G,G_1,G_2,G_3$ as above with $F=\{G,G_1,G_2,G_3\}$, the graphs $G/\{G,G_1,G_2\}$, $G_3/F$, and $G/F$ are shown in Fig. \ref{fig_shrink} (left, middle, and right, respectively).

We apply the UV divergence index $\omega$ from (\ref{eq_uv_div_index}) to the sets $s$ associated not only with $G$, but also with subgraphs $G'$ of $G$ or with graphs obtained from them by operations as described above. We denote it by $\omega_{G'}(s)$. Here $s\subseteq\edge[G]$; this means that we really apply it to $s\cap \edge[G']$. For instance, for $G,G_1,G_2,G_3$ as above with $F=\{G,G_1,G_2,G_3\}$, we have
\begin{gather*}
\omega_G(\{1,5,14\})=-2.5,\\ 
\omega_{G/F}(\{1,5,14\})=\omega_{G/F}(\{5,14\}) = -0.5,\\ 
\omega_{G_3/F}(\{1,5,14\}) = \omega_{G_3/F}(\emptyset) = 0,
\end{gather*}
\begin{gather*}
\omega_G(\{5,11,13\})=-3,\\ 
\omega_{G/F}(\{5,11,13\})=\omega_{G/F}(\{5\}) = -1,\\ 
\omega_{G_3/F}(\{5,11,13\}) = \omega_{G_3/F}(\{11,13\}) = -0.5.
\end{gather*}

\begin{figure*}
\includegraphics[width=150mm]{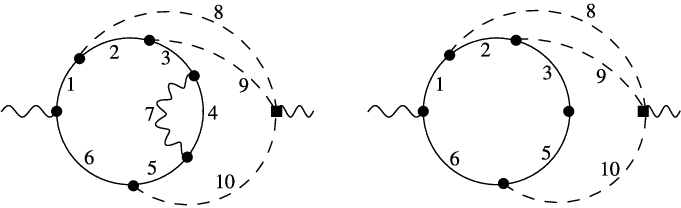}
\caption{Unfolded Feynman graph of (5) from Fig. \ref{fig23} (left), the one obtained from it by shrinking the fermion self-energy subgraph to a point (right).
\label{fig_with_fse}
}
\end{figure*}

There is a reason to change the definition of the UV divergence index for fermion self-energy graphs: since the mass part is taken on the mass shell (\ref{eq_operator_fse}) in the forest formula (\ref{eq_forest_formula}), this requires a different power counting in general ~\cite{volkov_iclos_npb}.
Suppose $G'$ is a graph obtained from $G$ by some operations as described above. Taking into account that all fermion self-energy graphs arising in $\widetilde{V}_{2j}$ contain only one loop, we define
$$
\overline{\omega}_{G'}(s)=\begin{cases} -|\photon(s)|(1-|\fermion(s)|), \\
\quad\quad\quad\text{ if $G'$ is fermion self-energy}, \\
\omega_{G'}(s)\text{ otherwise},
\end{cases}
$$
where $\photon(s)$ is the set of all photon lines in $s$. For example, if $G$ is the graph in Fig. \ref{fig_with_fse},left, and $G'=\langle 4,7\rangle$, then
$$
\overline{\omega}_G(\{3,4\})=\omega_G(\{3,4\})=-1,
$$
$$
\overline{\omega}_{G'}(\{3,4\})=\overline{\omega}_{G'}(\{4\})=0.
$$

A vertex $v$ is called \emph{intermediate} in a graph $G'$, if exactly two lines (internal or external) of $G'$ are incident to $v$. We say that a line $l\in\edge[G']$ is \emph{chained} in $G'$ if it is incident to some intermediate vertex of $G'$. For example, in the graph $G'$ of Fig. \ref{fig_with_fse},right there is a single intermediate vertex (the one connecting 3 and 5); the chained lines are $3$ and $5$. In the case of $\widetilde{V}_{2j}$, there can be at most one intermediate vertex and at most two chained lines. By definition,
$$
\ffch_{G'}(s) = \begin{cases} -0.5,\text{ if $G'$ has fermion chained lines,} \\
\quad\quad\quad\text{ and all of them are in $s$}, \\
-1,\text{ if $G'$ has photon or Coulomb chained} \\
\quad\quad\quad\text{ lines, and all of them are in $s$}, \\
0\text{ otherwise.}
\end{cases}
$$

The function $\tilde{\omega}_{G'}(s)$ is defined by
$$
\tilde{\omega}_{G'}(s)=\min(0,\ \overline{\omega}_{G'}(s)-\ffch_{G'}(s)).
$$
For instance, for the graph $G$ in Fig. \ref{fig_with_fse},right, we have
\begin{gather*}
\tilde{\omega}_{G}(\{1,2,3,6\}) = \overline{\omega}_{G}(\{1,2,3,6\}) = -1,\\ 
\tilde{\omega}_{G}(\{1,2,3,5,6\}) = \overline{\omega}_{G}(\{1,2,3,5,6\})+0.5 = 0.
\end{gather*}

Let us define the function $D_F(s)$ for $s\subseteq \edge[G]$ and a forest $F$ of subgraphs $G$:
$$
D_F(s) = -\sum_{G'\in F} \tilde{\omega}_{G'/F}(s).
$$
For example, if $G$ is the graph from Fig. \ref{fig_with_fse} (left), $G_1=\langle 4,7\rangle$ from the same figure, and $F=\{G,G_1\}$, then $G/F$ is exactly the graph from \ref{fig_with_fse} (right), and
\begin{gather*}
D_F(\{1,2,3,4,5,6\}) = -\tilde{\omega}_{G_1}(\{4\})-\tilde{\omega}_{G/F}(\{1,2,3,5,6\}) \\ = 0+0 = 0,
\end{gather*}
\begin{gather*}
D_F(\{1,2,3,6,7\}) = -\tilde{\omega}_{G_1}(\{7\})-\tilde{\omega}_{G/F}(\{1,2,3,6\})  \\ = 1+1 = 2.
\end{gather*}

We now introduce the function $\triangle_F(s)$, where $F$ is a forest of subgraphs of $G$ and $s\subseteq\edge[G]$, in order to handle the case $\vsp{p}\rightarrow 0$. Since $\vsp{p}\neq 0$, the divergence described in Sec. \ref{subsec_zero_momentum} does not occur. However, some of the integrals calculated tend to $\infty$ as $\vsp{p}\rightarrow 0$, and this behavior appears only in certain regions of the integration domains; it may therefore lead to Monte Carlo convergence rate issues. Put $\triangle_F(s)=1$ if all of the following conditions are satisfied:
\begin{enumerate}[(i)]
\item\label{triangle_cond_ferm} $\fermion(\edge[G])\subseteq s$;
\item\label{triangle_cond_path} $s$ does not contain a path between the external lines in $G/F$;
\item\label{triangle_cond_full} there is no $G'\in F$ such that $\edge[G'/F]\subseteq s$ and all external lines of $G'$ lie outside $s$;
\item\label{triangle_cond_zerodeg} there is no $G'\in F$ such that $\tilde{\omega}_{G'/F}(s)=0$ and $0<|s\cap \edge[G'/F]|<|\edge[G'/F]|$;
\end{enumerate}
otherwise, put $\triangle_F(s)=0$. 

These conditions reproduce the reasoning of Sec. \ref{subsec_zero_momentum} that leads to the reduction of power of $\delta$ at $\vsp{p}=0$. The first two conditions capture this effect for a single term of the forest formula; the last two account for cases where cancellation between different terms compensate it\footnote{Since the goal is only to construct a probability density function for Monte Carlo integration, an exact analysis is not required.}. Let us consider some examples. If $G$ is the graph in Fig. \ref{fig_v13_v15} (right), then
$$
\triangle_{\{G\}}(\{1,2,3,4,5,6\}) = 1,\ \ \triangle_{\{G\}}(\{1,2,3,4,5,6,7\}) = 0;
$$
in the second case, condition (\ref{triangle_cond_path}) is violated. For $G$ from Fig. \ref{fig_v13_v15} (left), we obtain
$$
\triangle_{\{G\}}(\{1,2,3,4\})=0,\ \ \triangle_{\{G,\langle 1,2,3,4\rangle\}}(\{1,2,3,4\})=0;
$$
in the first case, condition (\ref{triangle_cond_zerodeg}) is violated; in the second, condition (\ref{triangle_cond_full}) is violated. For $G$ from Fig. \ref{figdisclose} with $s=\{1,2,3,4,5,6,7,8,9\}$, we have
\begin{gather*}
\triangle_{\{G\}}(s)=1,\quad \triangle_{\{G,\langle 1,2,3,4\rangle\}}(s)=1,\\  \triangle_{\{G,\langle 6,7,8,9,11,12,13 \rangle \}}(s)=0;
\end{gather*}
in the first case, $\tilde{\omega}_G(s)=-1<0$ prevents violation of (\ref{triangle_cond_zerodeg}); in the second, line $5$ prevents violation of (\ref{triangle_cond_full}) for $G'=\langle 1,2,3,4\rangle$; in the third, (\ref{triangle_cond_path}) and (\ref{triangle_cond_zerodeg}) are violated. 

By $\forestsdeg[G]$ we denote the set of all forests consisting of self-energy, vertexlike, photon-photon scattering, or potential subgraphs of $G$ and satisfying the following conditions:
\begin{itemize}
\item $G\in F$;
\item $F$ contains all self-energy subgraphs of $G$.
\end{itemize}
For example, for $G$ in Fig. \ref{fig_with_fse} (left), we have
$$
\forestsdeg[G]=\{\{\langle 4,7 \rangle,G\},\ \{\langle 4,7 \rangle, \langle 1,2,3,4,5,6,7\rangle, G\}\}.
$$

By definition, set
\begin{equation}\label{eq_degz}
\ffdegz(s,\vsp{p}) = 
\begin{cases} 
\taumin(\czeromul/(-\log|\vsp{p}/m|),\czerosat), \\
\quad\quad\quad\text{if $|\vsp{p}/m|<1$} \\ 
\quad\quad\quad \text{and $\fermion(\edge[G])\subseteq s$ and} \\
\quad\quad\quad \text{there exists $F\in\forestsdeg[G]$ such that} \\
\quad\quad\quad\quad\quad D_F(s)-\triangle_F(s)\leq 0, \\
\taumin(\cinfmul/\log|\vsp{p}/m|,\cinfsat),\text{ if $|\vsp{p}/m|> 1$} \\
\quad\quad\quad\text{and $\fermion(\edge[G])\nsubseteq s$ and } \\
\quad\quad\quad\text{there exists $F\in\forestsdeg[G]$ such that} \\
\quad\quad\quad\quad\quad \text{$D_F(s)=0$} \\
\quad\quad\quad\quad\quad\text{and $s$ has a path between} \\
\quad\quad\quad\quad\quad\text{external lines in $G/F$}, \\
\taumax(\min_{F\in\forestsdeg[G]} [D_F(s)-\triangle_F(s)],\conesat) \\
\quad\quad\quad \text{ otherwise};
\end{cases}
\end{equation}
here
\begin{gather*}
\taumax(x,a) = \frac{a}{2}\mu((x-a)/a),\quad \mu(t)=2+t+\sqrt{t^2+0.25},\\ \taumin(x,a) = a\left( \frac{2}{1+e^{-2x/a}}-1 \right);
\end{gather*}
the functions $\taumax(x,a)$ and $\taumin(x,a)$ are smooth approximations of $\max(x,a)$ and $\min(x,a)$, respectively; the constants $\czerosat$, $\czeromul$, $\conesat$, $\cinfsat$, and $\cinfmul$ are arbitrary positive numbers that should be tuned to optimize Monte Carlo convergence.

The first condition in (\ref{eq_degz}) should work in the case when a divergence at $\vsp{p}=0$ occurs as described in Sec. \ref{subsec_zero_momentum}; in this case, $\ffdegz(s,\vsp{p})\rightarrow 0$ as $\vsp{p}\rightarrow 0$. The second condition of (\ref{eq_degz}) addresses an analogous convergence problem at $\vsp{p}\rightarrow\infty$; this situation may arise if replacing of $A(z)\vsp{p}^2+B(z)$ by $A(z)\vsp{p}^2$ in the denominators of (\ref{eq_integrand}) leads to divergence. The last condition in (\ref{eq_degz}) covers the ordinary case, where no dependence on $\vsp{p}$ is required. 

In practice, the vast majority of sets $s$ fall under the last condition of (\ref{eq_degz}). Most Feynman graphs have only one or two sets $s$ requiring $\vsp{p}$-dependent conditions. Only a small fraction of Hepp sectors are affected by them, yet these conditions significantly improve Monte Carlo convergence at very small and very large $|\vsp{p}|$.

Let us give some examples. Suppose $G$ is the graph from Fig. \ref{figdisclose}. Then
\begin{gather*}
\ffdegz(\{1,2,3,4,5,6,7,8,9\},\vsp{p}) \\
=\taumin(\czeromul/(-\log|\vsp{p}/m|),\czerosat)
\end{gather*}
for $|\vsp{p}/m|<1$, since this set contains all fermion lines, and for $F=\{G\}$ we have $D_F(\{1,2,3,4,5,6,7,8,9\})=1$ and $\triangle_F(\{1,2,3,4,5,6,7,8,9\})=1$.
On the other hand,
$$
\ffdegz(\{1,2,3,4,5\},\vsp{p})=\taumax(1,\conesat),
$$
since it does not contain all fermion lines, and $D_F(\{1,2,3,4,5\})=1$, $\triangle_F(\{1,2,3,4,5\})=0$ for all $F$. We also have
$$
\ffdegz(\{5,10,14\},\vsp{p}) = \taumin(\cinfmul/\log|\vsp{p}/m|,\cinfsat)
$$
for $|\vsp{p}/m|>1$, because for 
$$
F=\{G,\langle 1,2,3,4\rangle, \langle 6,7,8,9,11,12,13\rangle\}
$$ 
we obtain $D_F(\{5,10,14\})=0$, and $G/F$ contains a path  through $\{5,10,14\}$ connecting the external lines. Also,
$$
\ffdegz(\{2,4,10,14\},\vsp{p})=\taumax(1.5,\conesat),
$$ 
since $D_F(\{2,4,10,14\})=1.5$ for $F=\{G,\langle 1,2,3,4\rangle \}$. 

We use the parameter values
\begin{equation}\label{eq_degz_constants}
\begin{array}{c}
\czerosat=0.9,\quad \czeromul=3.75,\quad \conesat=0.5,\\ 
\cinfsat=0.5,\quad \cinfmul=3
\end{array}
\end{equation}
throughout all our calculations. Note that the values of $\czeromul$ and $\cinfmul$ are relatively large: a probability density function constructed with smaller values would better approximate the integrand. However, too small values of $\czeromul$ and $\cinfmul$ slow down the computation, as many samples would then require high-precision arithmetic (see Sec. \ref{subsec_machine}). Thus, the chosen larger values represent a compromise, while additional stability term constructed with smaller values is also employed (see Sec. \ref{subsec_stability}).

\subsection{Improving the stability}\label{subsec_stability}

To make Monte Carlo convergence as fast as possible, one should minimize
$$
S=\int_{z>0} \frac{F(\vsp{p},z)^2}{g(z,\vsp{p})}\delta(z_1+\ldots+z_n-1) d^n z,
$$
where $F$ is the integrand, and $g$ is the probability density function. The  following situations should be avoided:
\begin{enumerate}[(i)]
\item\label{item_infinite_var} $S$ is infinite (an underestimation of the asymptotic behavior of $|F(\vsp{p},z)|$ in even a single Hepp sector may lead to this, even if the approximation works well in all other sectors; for example, for $\widetilde{V}_{27}$ there are $2^{18}=262144$ sets and $18!=6402373705728000$ sectors for each Feynman graph, all of which must be controlled);
\item\label{item_rarely_visited} a region of the integration domain contributes significantly, but is visited only rarely during the integration;
\item\label{item_big_early} extremely large values of $|F(\vsp{p},z)/g(z,\vsp{p})|$ occur too early in the integration process and spoil the result (this is possible because $F(\vsp{p},z)/g(z,\vsp{p})$ is generally unbounded);
\item\label{item_error_underest} the integration error is underestimated due to unvisited regions with large $|F(\vsp{p},z)/g(z,\vsp{p})|$.
\end{enumerate}

To handle (\ref{item_infinite_var}) and (\ref{item_rarely_visited}), we introduce stabilization terms into the probability density function:
\begin{gather*}
g(z,\vsp{p}) = C_{\mcmain}\times g_0(z,\vsp{p})+C_{\mcmodify}\times g_1(z,\vsp{p}) + \\ C_{\mcsector}\times g_{\mcsector}(z,\vsp{p})+C_{\mcuniform}\times g_{\mcuniform}(z) + \\ C_{\mcmin}\times g_{\mcmin}(z);
\end{gather*}
here $g_0$ is constructed by the rules of Sec. \ref{subsec_degz} with (\ref{eq_degz_constants}); $g_1$ is constructed with the same rules but with
$$
\czeromul=\cinfmul=1;
$$
$g_{\mcsector}$ is defined by the rules of Sec. \ref{subsec_degz} with (\ref{eq_degz_constants}), but normalized differently in different sectors so that the integral of $g_{\mcsector}$ over each Hepp's sector is the same;
$g_{\mcmin}$ is constructed with $\ffdeg(s,\vsp{p})=R$, where $R>0$ is a constant; $g_{\mcuniform}$ is a constant function normalized so that (\ref{eq_dens_normalization}) holds for it instead of $g_0$; The coefficients $C_{\mcmain}$, $C_{\mcmodify}$, $C_{\mcsector}$, $C_{\mcuniform}$, and $C_{\mcmin}$ are arbitrary nonnegative constants satisfying
$$
C_{\mcmain}+C_{\mcmodify}+C_{\mcsector}+C_{\mcuniform}+C_{\mcmin}=1.
$$
In our calculations we use
\begin{gather*}
C_{\mcsector}=0.05,\quad C_{\mcuniform}=0.06,\\ 
C_{\mcmin}=0.04,\quad R=0.75, \\
C_{\mcmodify}=0.005\text{ for }V_{23},\quad C_{\mcmodify}=0.03\text{ for }V_{25},V_{27}.
\end{gather*}
In many cases, $g_1$ approximates the integrand behavior better than $g_0$ as $\vsp{p}\rightarrow 0,\infty$; however, using $g_1$ as a probability density function slows down computations due to high-precision arithmetic (see \ref{subsec_machine}). This effect is compensated by the smallness of $C_{\mcmodify}$.

To mitigate the situation (\ref{item_big_early}), we use the techniques described in Section III.D of \cite{volkov_prd}, adapted for GPUs. To address (\ref{item_error_underest}), we employ the heuristic from Section IV.F of \cite{volkov_gpu}.

\subsection{Machine implementation}\label{subsec_machine}

\begin{table*}
\begin{center}
\caption{Contributions of the Monte-Carlo samples with round-off errors for several values of $\vsp{p}$; the momentum $\vsp{p}$ is given in units of $m$; the factor $m^{-2}$ is omitted.
\label{table_round_off_contributions}
}
\begin{ruledtabular}
\begin{tabular}{cccccc}
$|\vsp{p}|$ & Value & $\triangle^{\text{fail}}_{\text{EIA}}$ & $\triangle^{\text{fail}}_{\text{IA}}$ & $\triangle^{\text{fail}}_{\text{128}}$ & $\triangle^{\text{fail}}_{\text{256}}$ \\
\hline 
\multicolumn{6}{c}{Evaluation of $V_{23}$} \\
\hline 
$0.001$ & $0.062303(73)$ & $0.32$ & $0.5$ & $0.0039$ & $-5.6\times 10^{-8}$ \\
$1$ & $0.044156(40)$ & $0.013$ & $0.00069$ & $-0.00001$ & $0$ \\
$1000$ & $0.0000013361(10)$ & $0.0000011$ & $2.6\times 10^{-7}$ & $1.1\times 10^{-9}$ & $0$ \\
$100000$ & $2.3622(18)\times 10^{-10}$ & $1.1\times 10^{-11}$ & $9.2\times 10^{-11}$ & $4.5\times 10^{-13}$ & $-9.5\times 10^{-22}$ \\
\hline 
\multicolumn{6}{c}{Evaluation of $V_{25}$} \\
\hline 
$0.001$ & $0.014832(67)$ & $0.044$ & $0.023$ & $0.0021$ & $-1.6\times 10^{-15}$ \\
$1$ & $0.012235(52)$ & $0.013$ & $0.0012$ & $6\times 10^{-6}$ & $0$ \\
$1000$ & $6.649(25)\times 10^{-7}$ & $8.7\times 10^{-7}$ & $-7.2\times 10^{-8}$ & $-3.5\times 10^{-10}$ & $-6.3\times 10^{-69}$ \\
$100000$ & $1.1819(44)\times 10^{-10}$ & $3.5\times 10^{-10}$ & $-2.8\times 10^{-11}$ & $-1.2\times 10^{-13}$ & $-1.4\times 10^{-65}$ \\
\hline 
\multicolumn{6}{c}{Evaluation of $V_{27}$} \\
\hline 
$0.001$ & $0.00824(14)$ & $0.0043$ & $-0.00097$ & $-7.3\times 10^{-6}$ & $-6.1\times 10^{-22}$ \\
$1$ & $0.00708(12)$ & $0.0032$ & $-0.00089$ & $-6.4\times 10^{-6}$ & $0$ \\
$1000$ & $4.596(70)\times 10^{-7}$ & $3.8\times 10^{-7}$ & $-6.3\times 10^{-8}$ & $-4.9\times 10^{-10}$ & $2.5\times 10^{-67}$ \\
$100000$ & $8.10(12)\times 10^{-11}$ & $7.2\times 10^{-11}$ & $-1.8\times 10^{-11}$ & $-1.3\times 10^{-13}$ & $-2.2\times 10^{-67}$
\end{tabular}
\end{ruledtabular}
\end{center}
\end{table*}

\begin{table*}
\begin{center}
\caption{Statistics of round-off errors in the integrand evaluations for several values of $\vsp{p}$; the momentum $\vsp{p}$ is given in units of $m$.
\label{table_round_off_numbers}
}
\begin{ruledtabular}
\begin{tabular}{ccccccc}
$|\vsp{p}|$ & $N_{\text{total}}$ & $N^{\text{fail}}_{\text{EIA}}$ & $N^{\text{fail}}_{\text{IA}}$ & $N^{\text{fail}}_{\text{128}}$ & $N^{\text{fail}}_{\text{256}}$ & $N^{\text{fail}}_{\text{384}}$ \\
\hline 
\multicolumn{7}{c}{Evaluation of $V_{23}$} \\
\hline 
$0.001$ & $6\times 10^{11}$ & $71\times 10^9$ & $47\times 10^9$ & $29\times 10^8$ & $44\times 10^6$ & $34\times 10^5$ \\
$1$ & $21\times 10^9$ & $12\times 10^8$ & $23\times 10^7$ & $5\times 10^5$ & $0$ & $0$ \\
$1000$ & $15\times 10^{10}$ & $13\times 10^9$ & $27\times 10^8$ & $72\times 10^5$ & $0$ & $0$ \\
$100000$ & $5\times 10^{11}$ & $82\times 10^9$ & $75\times 10^8$ & $21\times 10^6$ & $2$ & $0$ \\
\hline 
\multicolumn{7}{c}{Evaluation of $V_{25}$} \\
\hline 
$0.001$ & $7\times 10^{10}$ & $12\times 10^9$ & $31\times 10^8$ & $2\times 10^8$ & $67\times 10^5$ & $45\times 10^4$ \\
$1$ & $11\times 10^9$ & $17\times 10^8$ & $12\times 10^7$ & $22\times 10^4$ & $0$ & $0$ \\
$1000$ & $33\times 10^{9}$ & $7\times 10^9$ & $49\times 10^7$ & $14\times 10^5$ & $569$ & $1$ \\
$100000$ & $87\times 10^{9}$ & $23\times 10^9$ & $18\times 10^8$ & $11\times 10^6$ & $15\times 10^4$ & $3232$ \\
\hline 
\multicolumn{7}{c}{Evaluation of $V_{27}$} \\
\hline 
$0.001$ & $42\times 10^8$ & $15\times 10^8$ & $46\times 10^6$ & $16\times 10^5$ & $33870$ & $1456$ \\
$1$ & $4\times 10^9$ & $14\times 10^8$ & $27\times 10^6$ & $49427$ & $0$ & $0$ \\
$1000$ & $82\times 10^8$ & $36\times 10^8$ & $84\times 10^6$ & $288675$ & $200$ & $0$ \\
$100000$ & $19\times 10^9$ & $87\times 10^8$ & $4\times 10^8$ & $29\times 10^5$ & $44360$ & $941$
\end{tabular}
\end{ruledtabular}
\end{center}
\end{table*}

The most resource-intensive part of the computation is the Monte Carlo integration. We employed NVidia P100 GPUs (each of them was accompanied by one core of the CPU Intel Xeon E5-2698). The integration required 15 GPU-days for each of $V_{23},V_{25}$ and 42 GPU-days for $V_{27}$. To obtain the results presented in ~\cite{vp_coulomb_potential_2025}, two random number generators from the NVidia CURAND library were used: \texttt{MRG32k3a} and \texttt{Philox\char`_4x32\char`_10}. The results from different generators were compared and combined; no statistically significant difference between them was detected.

The integration algorithm is nonadaptive within a single integral. However, the program evaluated integrals for all Feynman graphs simultaneously, choosing the graph number randomly after each block of samples. This graph selection algorithm is adaptive and explained in ~\cite{volkov_5loops_2019}.

In our approach, all divergences are cancelled numerically at the integrand level. This cancellation introduces round-off errors. To control them, integrand values are evaluated using interval arithmetic. Samples producing overly large intervals are recomputed with higher precision. The acceptance-rejection algorithm for intervals  is described in detail in Section IV.A of \cite{volkov_5loops_2019}. It ensures that the round-off error remains small relative to the Monte Carlo statistical error even in the worst case and accounts for the fact that round-off errors may have a nonzero mean and thus cannot simply be added as statistical errors.

The compiled integrand code sizes are 30 MB for $V_{23}$, 450 MB for $V_{25}$, and 11 GB for $V_{27}$. Special techniques are required to evaluate such huge functions on GPUs. The general scheme of the implementation, as well as the interval arithmetic, are described in \cite{volkov_5loops_2019}. We generated the integrand code separately for different precisions:
\begin{itemize}
\item ``Eliminated Interval Arithmetic'' (EIA) is an approach that uses machine \texttt{double} precision, but is significantly faster than standard interval arithmetic (at the cost of larger intervals). It is described in Section IV.C of \cite{volkov_gpu}.
\item Standard machine \texttt{double}-precision interval arithmetic.
\item Arbitrary-precision interval arithmetic with a variable mantissa size (128-bit, 256-bit, and 384-bit mantissas were used in our calculations).
\end{itemize}

The code generator for the integrands was written in the \textsc{d} programming language. The generated code was in \textsc{c++}\footnote{Formally \textsc{c++}, but mostly using \textsc{c}-style constructs.} with \textsc{cuda}.

Table \ref{table_round_off_contributions} reports the contributions of Monte Carlo samples requiring increased precision, for four values of $|\vsp{p}|$. $\triangle^{\text{fail}}_{\text{EIA}}$, $\triangle^{\text{fail}}_{\text{IA}}$, $\triangle^{\text{fail}}_{\text{128}}$, and $\triangle^{\text{fail}}_{\text{256}}$ denote contributions from samples where the following precisions failed, respectively: EIA, standard \texttt{double}-precision interval arithmetic, 128-bit-mantissa, and 256-bit-mantissa. The table shows that high-precision arithmetic is especially important for $\vsp{p}$ near $0$.

The corresponding numbers of Monte Carlo samples are given in Table \ref{table_round_off_numbers}. The table also includes $N^{\text{fail}}_{\text{384}}$, the number of the samples for which 384-bit-mantissa precision failed. Since this is the highest precision used in our integration, such samples are treated as contributing $0$.

\section{Results for individual Feynman graphs}\label{sec_individual_results}

The values of individual graph are essential for verifying the results. They are presented in Tables \ref{table_v23_individual}, \ref{table_v25_individual}, and \ref{table_v27_individual} for four values of $|\vsp{p}|$. Some of these values are independent of $(m_0)^2$ from (\ref{eq_operator_vertex}) and (\ref{eq_operator_fse}). The $(m_0)^2$-dependent values are given for $(m_0)^2=m^2$, $(m_0)^2=0$, $(m_0)^2=-m^2$, $(m_0)^2=-5m^2$, and $(m_0)^2=-50m^2$. Integrals of the form
$$
\int_{z>0} |F_j(\vsp{p},z)|\delta(z_1+\ldots+z_n-1)d^n z, 
$$
where $F_j$ is the integrand of the $j$-th graph, illustrate the degree of oscillations in the integrand. The corresponding sums are also provided in the tables. Each summation runs over the graphs of the current block; graph 1 is excluded, since its contribution is obtained from lower-order values. The subtraction point $(m_0)^2=-5m^2$ is optimal with respect to the degree of oscillations inside the integrals for $V_{23}$ and $V_{25}$; it was therefore used to obtain the final results. The oscillations tend to $\infty$ as $\vsp{p}\rightarrow 0$, which is noticeable in the tables. The independence of the final results of $(m_0)^2$ is also shown in the tables. The values for different $(m_0)^2$ were obtained by direct calculation; however, it is worth noting that these values could also be determined more precisely from one-loop ones.

\begin{longtable*}{ccccc}\caption{Contributions of individual Feynman graphs to $\widetilde{V}_{23}(\vsp{p})$ for several values of $\vsp{p}$ and $(m_0)^2$ from (\ref{eq_operator_vertex}) and (\ref{eq_operator_fse}).  The column ``Object'' refers to the graph number from Fig.~\ref{fig23} or to a special entry. The integral sums do not include graph 1. The momentum $\vsp{p}$ is given in units of $m$; the factor $m^{-2}$ is omitted. (continued)} \\
\hline \hline Object & $|\vsp{p}|=0.001$ & $|\vsp{p}|=1$ & $|\vsp{p}|=1000$ & $|\vsp{p}|=100000$ \\ \hline  \endhead
\caption{Contributions of individual Feynman graphs to $\widetilde{V}_{23}(\vsp{p})$ for several values of $\vsp{p}$ and $(m_0)^2$ from (\ref{eq_operator_vertex}) and (\ref{eq_operator_fse}).  The column ``Object'' refers to the graph number from Fig.~\ref{fig23} or to a special entry. The integral sums do not include graph 1. The momentum $\vsp{p}$ is given in units of $m$; the factor $m^{-2}$ is omitted.}\label{table_v23_individual} \\
\hline \hline Object & $|\vsp{p}|=0.001$ & $|\vsp{p}|=1$ & $|\vsp{p}|=1000$ & $|\vsp{p}|=100000$ \\ \hline  \endfirsthead
\hline \endfoot  \hline \hline \endlastfoot
Value & $0.062303(73)$ & $0.044156(40)$ & $0.0000013361(10)$ & $2.3622(18)\times 10^{-10}$ \\
\hline
\multicolumn{5}{c}{Contributions independent of $(m_0)^2$} \\
\hline 
1 & $1.20249(41)\times 10^{-9}$ & $0.00082412(31)$ & $3.3938(13)\times 10^{-7}$ & $5.9656(13)\times 10^{-11}$ \\
2 & $0.0371916(31)$ & $0.0335023(79)$ & $1.99164(36)\times 10^{-6}$ & $3.45914(65)\times 10^{-10}$ \\
3 & $-0.0148240(20)$ & $-0.0134644(46) $ & $-9.0500(25)\times 10^{-7}$ & $-1.60150(46)\times 10^{-10}$ \\
9 & $-1.390928(24)$ & $-0.125532(17)$ & $2.565698(37)\times 10^{-5}$ & $7.656663(65)\times 10^{-9}$ \\
$\sum_j \int |F_j(z)|dz$ & $2.263$ & $0.44$ & $4.89\times 10^{-5}$ & $1.44\times 10^{-8}$ \\
\hline 
\multicolumn{5}{c}{Dependent on $(m_0)^2$ contributions, $(m_0)^2=m^2$} \\
\hline 
4 & $0.34568(25)$ & $0.087764(29)$ & $4.5129(35)\times 10^{-6}$ & $1.2901(12)\times 10^{-9}$ \\
5 & $-0.67087(28)$ & $-0.057568(28)$ & $4.6666(39)\times 10^{-6}$ & $1.34215(70)\times 10^{-9}$ \\
6 & $1.01223(23)$ & $0.059736(24)$ & $-1.77582(37)\times 10^{-5}$ & $-5.1984(11)\times 10^{-9}$ \\
7 & $0.59275(18)$ & $0.073429(16)$ & $-8.0111(25)\times 10^{-6}$ & $-2.45701(73)\times 10^{-9}$ \\
8 & $0.15126(15)$ & $-0.014457(17)$ & $-9.1512(19)\times 10^{-6}$ & $-2.64145(54)\times 10^{-9}$ \\
$\sum_j \int F_j(z)dz$ & $1.43105(50)$ & $0.148904(52)$ & $-2.57409(71)\times 10^{-5}$ & $-7.6647(20)\times 10^{-9}$ \\
$\sum_j \int |F_j(z)|dz$ & $8.128$ & $1.54$ & $7.489\times 10^{-5}$ & $1.93\times 10^{-8}$ \\
\hline 
\multicolumn{5}{c}{Dependent on $(m_0)^2$ contributions, $(m_0)^2=0$} \\
\hline 
4 & $0.33205(19)$ & $0.077526(22)$ & $4.4440(33)\times 10^{-6}$ & $1.2839(10)\times 10^{-9}$ \\
5 & $-0.68485(22)$ & $-0.067785(21)$ & $4.6011(31)\times 10^{-6}$ & $1.33650(78)\times 10^{-9}$ \\
6 & $1.02610(19)$ & $0.069930(19)$ & $-1.76972(32)\times 10^{-5}$ & $-5.1922(10)\times 10^{-9}$ \\
7 & $0.59956(13)$ & $0.078518(12)$ & $-7.9766(22)\times 10^{-6}$ & $-2.45508(69)\times 10^{-9}$ \\
8 & $0.15805(11)$ & $-0.009328(12)$ & $-9.1210(22)\times 10^{-6}$ & $-2.63863(42)\times 10^{-9}$ \\
$\sum_j \int F_j(z)dz$ & $1.43090(39)$ & $0.148861(40)$ & $-2.57497(64)\times 10^{-5}$ & $-7.6655(18)\times 10^{-9}$ \\
$\sum_j \int |F_j(z)|dz$ & $5.708$ & $1.073$ & $6.638\times 10^{-5}$ & $1.791\times 10^{-8}$ \\
\hline 
\multicolumn{5}{c}{Dependent on $(m_0)^2$ contributions, $(m_0)^2=-m^2$} \\
\hline 
4 & $0.32724(16)$ & $0.074134(19)$ & $4.4218(31)\times 10^{-6}$ & $1.2826(10)\times 10^{-9}$ \\
5 & $-0.68904(19)$ & $-0.071242(19)$ & $4.5808(29)\times 10^{-6}$ & $1.33340(90)\times 10^{-9}$ \\
6 & $1.03013(16)$ & $0.073339(17)$ & $-1.76744(23)\times 10^{-5}$ & $-5.1902(10)\times 10^{-9}$ \\
7 & $0.60163(12)$ & $0.080211(12)$ & $-7.9645(21)\times 10^{-6}$ & $-2.45388(69)\times 10^{-9}$ \\
8 & $0.16038(10)$ & $-0.007605(11)$ & $-9.1135(22)\times 10^{-6}$ & $-2.63679(69)\times 10^{-9}$ \\
$\sum_j \int F_j(z)dz$ & $1.43035(34)$ & $0.148837(36)$ & $-2.57498(56)\times 10^{-5}$ & $-7.6649(19)\times 10^{-9}$ \\
$\sum_j \int |F_j(z)|dz$ & $5.104$ & $0.939$ & $6.42\times 10^{-5}$ & $1.756\times 10^{-8}$ \\
\hline 
\multicolumn{5}{c}{Dependent on $(m_0)^2$ contributions, $(m_0)^2=-5m^2$} \\
\hline 
4 & $0.319246(29)$ & $0.067837(16)$ & $4.38801(42)\times 10^{-6}$ & $1.277854(76)\times 10^{-9}$ \\
5 & $-0.697473(36)$ & $-0.077495(18)$ & $4.54248(41)\times 10^{-6}$ & $1.329490(72)\times 10^{-9}$ \\
6 & $1.038674(37)$ & $0.079617(17)$ & $-1.763637(47)\times 10^{-5}$ & $-5.186314(84)\times 10^{-9}$ \\
7 & $0.605919(26)$ & $0.083356(12)$ & $-7.94803(31)\times 10^{-6}$ & $-2.451390(55)\times 10^{-9}$ \\
8 & $0.164498(22)$ & $-0.004489(12)$ & $-9.09303(32)\times 10^{-6}$ & $-2.635494(58)\times 10^{-9}$ \\
$\sum_j \int F_j(z)dz$ & $1.430864(68)$ & $0.148826(34)$ & $-2.574693(87)\times 10^{-5}$ & $-7.66585(16)\times 10^{-9}$ \\
$\sum_j \int |F_j(z)|dz$ & $4.495$ & $0.756$ & $6.162\times 10^{-5}$ & $1.718\times 10^{-8}$ \\
\hline 
\multicolumn{5}{c}{Dependent on $(m_0)^2$ contributions, $(m_0)^2=-50m^2$} \\
\hline 
4 & $0.30101(16)$ & $0.054011(20)$ & $4.3017(26)\times 10^{-6}$ & $1.2693(10)\times 10^{-9}$ \\
5 & $-0.71598(20)$ & $-0.091331(20)$ & $4.4587(18)\times 10^{-6}$ & $1.32007(56)\times 10^{-9}$ \\
6 & $1.05688(22)$ & $0.093448(22)$ & $-1.75516(40)\times 10^{-5}$ & $-5.1749(10)\times 10^{-9}$ \\
7 & $0.61526(16)$ & $0.090312(16)$ & $-7.9060(14)\times 10^{-6}$ & $-2.44720(66)\times 10^{-9}$ \\
8 & $0.17375(14)$ & $0.002440(14)$ & $-9.0630(66)\times 10^{-6}$ & $-2.63166(67)\times 10^{-9}$ \\
$\sum_j \int F_j(z)dz$ & $1.43091(40)$ & $0.148880(41)$ & $-2.57601(85)\times 10^{-5}$ & $-7.6643(18)\times 10^{-9}$ \\
$\sum_j \int |F_j(z)|dz$ & $6.364$ & $0.956$ & $5.7\times 10^{-5}$ & $1.647\times 10^{-8}$ \\
\end{longtable*}

\begin{longtable*}{ccccc}\caption{Contributions of individual Feynman graphs to $\widetilde{V}_{25}(\vsp{p})$ for several values of $\vsp{p}$ and $(m_0)^2$ from (\ref{eq_operator_vertex}) and (\ref{eq_operator_fse}).  The column ``Object'' refers to the graph number from Fig.~\ref{fig25} or to a special entry. The integral sums do not include graph 1. The momentum $\vsp{p}$ is given in units of $m$; the factor $m^{-2}$ is omitted.(continued)} \\
\hline \hline Object & $|\vsp{p}|=0.001$ & $|\vsp{p}|=1$ & $|\vsp{p}|=1000$ & $|\vsp{p}|=100000$ \\ \hline  \endhead
\caption{Contributions of individual Feynman graphs to $\widetilde{V}_{25}(\vsp{p})$ for several values of $\vsp{p}$ and $(m_0)^2$ from (\ref{eq_operator_vertex}) and (\ref{eq_operator_fse}).  The column ``Object'' refers to the graph number from Fig.~\ref{fig25} or to a special entry. The integral sums do not include graph 1. The momentum $\vsp{p}$ is given in units of $m$; the factor $m^{-2}$ is omitted.}\label{table_v25_individual} \\
\hline \hline Object & $|\vsp{p}|=0.001$ & $|\vsp{p}|=1$ & $|\vsp{p}|=1000$ & $|\vsp{p}|=100000$ \\ \hline  \endfirsthead
\hline \endfoot  \hline \hline \endlastfoot
Value & $0.014832(67)$ & $0.12235(52)$ & $6.649(25)\times 10^{-7}$ & $1.1819(44)\times 10^{-10}$ \\
\hline
\multicolumn{5}{c}{Contributions independent of $(m_0)^2$} \\
\hline 
1 & $2.63353(81)\times 10^{-10}$ & $0.00019845(12)$ & $1.15490(73)\times 10^{-7}$ & $2.02959(13)\times 10^{-11}$ \\
2 & $0.0143827(39)$ & $0.0128682(80)$ & $7.1296(50)\times 10^{-7}$ & $1.22948(88)\times 10^{-10}$ \\
3 & $-0.0121048(38)$ & $-0.0109832(76)$ & $-7.2071(52)\times 10^{-7}$ & $-1.27559(93)\times 10^{-10}$ \\
4 & $0.0125484(29)$ & $0.0112766(58)$ & $6.4791(35)\times 10^{-7}$ & $1.12321(62)\times 10^{-10}$ \\
5 & $-0.003205(24)$ & $-0.002782(25)$ & $-2.42(14)\times 10^{-8}$ & $-2.74(27)\times 10^{-12}$ \\
6 & $0.001452(15)$ & $0.001243(16)$ & $1.051(95)\times 10^{-8}$ & $9.1(19)\times 10^{-13}$ \\
9 & $-0.281122(23)$ & $-0.085039(13)$ & $-1.65799(32)\times 10^{-6}$ & $-2.69632(41)\times 10^{-10}$ \\
10 & $0.364526(27)$ & $0.105882(14)$ & $2.13992(30)\times 10^{-6}$ & $3.48926(39)\times 10^{-10}$ \\
11 & $-0.277184(22)$ & $-0.082434(13)$ & $-1.67281(30)\times 10^{-6}$ & $-2.72656(39)\times 10^{-10}$ \\
15 & $-0.242679(23)$ & $-0.056145(12)$ & $-1.23912(28)\times 10^{-6}$ & $-2.03060(37)\times 10^{-10}$ \\
16 & $0.161084(19)$ & $0.0363764(87)$ & $7.5555(17)\times 10^{-7}$ & $1.23512(24)\times 10^{-10}$ \\
$\sum_j \int |F_j(z)|dz$ & $1.66$ & $0.652$ & $4.322\times 10^{-5}$ & $1.114\times 10^{-8}$ \\
\hline
\multicolumn{5}{c}{Dependent on $(m_0)^2$ contributions, $(m_0)^2=m^2$} \\
\hline 
7 & $0.00302(18)$ & $0.01106(10)$ & $-4.32(18)\times 10^{-8}$ & $-1.380(30)\times 10^{-11}$ \\
8 & $0.11062(20)$ & $0.031114(68)$ & $4.489(18)\times 10^{-7}$ & $7.546(30)\times 10^{-11}$ \\
12 & $0.065419(94)$ & $0.024123(37)$ & $4.696(11)\times 10^{-7}$ & $7.785(18)\times 10^{-11}$ \\
13 & $-0.03869(14)$ & $-0.019403(86)$ & $1.125(18)\times 10^{-7}$ & $2.277(31)\times 10^{-11}$ \\
14 & $0.08542(18)$ & $0.013553(65)$ & $4.226(17)\times 10^{-7}$ & $7.483(29)\times 10^{-11}$ \\
17 & $0.00897(13)$ & $0.015778(87)$ & $8.25(18)\times 10^{-8}$ & $7.85(31)\times 10^{-12}$ \\
18 & $0.04209(12)$ & $0.005487(45)$ & $1.053(12)\times 10^{-7}$ & $1.922(20)\times 10^{-11}$ \\
$\sum_j \int F_j(z)dz$ & $0.27685(41)$ & $0.08172(19)$ & $1.5981(43)\times 10^{-6}$ & $2.6417(73)\times 10^{-10}$ \\
$\sum_j \int |F_j(z)|dz$ & $1.168$ & $0.771$ & $5.109\times 10^{-5}$ & $1.232\times 10^{-8}$ \\
\hline
\multicolumn{5}{c}{Dependent on $(m_0)^2$ contributions, $(m_0)^2=0$} \\
\hline 
7 & $-0.00011(15)$ & $0.00870(10)$ & $-6.13(32)\times 10^{-8}$ & $-1.558(53)\times 10^{-11}$ \\
8 & $0.11376(44)$ & $0.03349(10)$ & $4.673(31)\times 10^{-7}$ & $7.713(53)\times 10^{-11}$ \\
12 & $0.06639(19)$ & $0.025243(45)$ & $4.792(19)\times 10^{-7}$ & $7.915(32)\times 10^{-11}$ \\
13 & $-0.04166(16)$ & $-0.02182(13)$ & $8.91(32)\times 10^{-8}$ & $2.050(54)\times 10^{-11}$ \\
14 & $0.08805(26)$ & $0.015925(76)$ & $4.483(31)\times 10^{-7}$ & $7.668(52)\times 10^{-11}$ \\
17 & $0.00603(17)$ & $0.01312(11)$ & $5.76(33)\times 10^{-8}$ & $5.33(55)\times 10^{-12}$ \\
18 & $0.04409(12)$ & $0.006642(42)$ & $1.157(21)\times 10^{-7}$ & $1.989(37)\times 10^{-11}$ \\
$\sum_j \int F_j(z)dz$ & $0.27656(62)$ & $0.08130(24)$ & $1.5960(77)\times 10^{-6}$ & $2.631(13)\times 10^{-10}$ \\
$\sum_j \int |F_j(z)|dz$ & $0.898$ & $0.551$ & $4.393\times 10^{-5}$ & $1.111\times 10^{-8}$ \\
\hline
\multicolumn{5}{c}{Dependent on $(m_0)^2$ contributions, $(m_0)^2=-m^2$} \\
\hline 
7 & $-0.00138(16)$ & $0.007972(80)$ & $-7.28(21)\times 10^{-8}$ & $-1.692(36)\times 10^{-11}$ \\
8 & $0.11515(21)$ & $0.034480(77)$ & $4.783(21)\times 10^{-7}$ & $7.853(36)\times 10^{-11}$ \\
12 & $0.067535(89)$ & $0.025779(45)$ & $4.824(13)\times 10^{-7}$ & $7.972(22)\times 10^{-11}$ \\
13 & $-0.04286(13)$ & $-0.022612(58)$ & $8.37(22)\times 10^{-8}$ & $1.975(36)\times 10^{-11}$ \\
14 & $0.08926(19)$ & $0.016822(72)$ & $4.526(21)\times 10^{-7}$ & $7.779(35)\times 10^{-11}$ \\
17 & $0.00496(15)$ & $0.012476(79)$ & $5.82(22)\times 10^{-8}$ & $3.96(37)\times 10^{-12}$ \\
18 & $0.04426(10)$ & $0.007074(34)$ & $1.220(15)\times 10^{-7}$ & $2.162(25)\times 10^{-11}$ \\
$\sum_j \int F_j(z)dz$ & $0.27694(40)$ & $0.08199(17)$ & $1.6045(52)\times 10^{-6}$ & $2.6444(87)\times 10^{-10}$ \\
$\sum_j \int |F_j(z)|dz$ & $0.823$ & $0.486$ & $4.155\times 10^{-5}$ & $1.071\times 10^{-8}$ \\
\hline
\multicolumn{5}{c}{Dependent on $(m_0)^2$ contributions, $(m_0)^2=-5m^2$} \\
\hline 
7 & $-0.0030012(89)$ & $0.006413(14)$ & $-8.506(58)\times 10^{-8}$ & $-1.791(10)\times 10^{-11}$ \\
8 & $0.116644(17)$ & $0.035940(13)$ & $4.8990(60)\times 10^{-7}$ & $7.997(10)\times 10^{-11}$ \\
12 & $0.068366(11)$ & $0.0265454(90)$ & $4.8871(35)\times 10^{-7}$ & $8.0123(60)\times 10^{-11}$ \\
13 & $-0.0445273(91)$ & $-0.024179(12)$ & $7.057(60)\times 10^{-8}$ & $1.828(10)\times 10^{-11}$ \\
14 & $0.091298(16)$ & $0.018311(12)$ & $4.6689(59)\times 10^{-7}$ & $7.907(10)\times 10^{-11}$ \\
17 & $0.0031859(93)$ & $0.010914(11)$ & $4.116(60)\times 10^{-8}$ & $3.88(10)\times 10^{-12}$ \\
18 & $0.045167(12)$ & $0.0078282(90)$ & $1.2521(42)\times 10^{-7}$ & $2.1516(70)\times 10^{-11}$ \\
$\sum_j \int F_j(z)dz$ & $0.277133(32)$ & $0.081773(31)$ & $1.5974(14)\times 10^{-6}$ & $2.6493(24)\times 10^{-10}$ \\
$\sum_j \int |F_j(z)|dz$ & $0.72$ & $0.391$ & $3.718\times 10^{-5}$ & $9.969\times 10^{-9}$ \\
\hline
\multicolumn{5}{c}{Dependent on $(m_0)^2$ contributions, $(m_0)^2=-50m^2$} \\
\hline 
7 & $-0.00685(14)$ & $0.003045(80)$ & $-1.147(17)\times 10^{-7}$ & $-2.092(29)\times 10^{-11}$ \\
8 & $0.12061(23)$ & $0.039181(65)$ & $5.192(18)\times 10^{-7}$ & $8.307(30)\times 10^{-11}$ \\
12 & $0.07042(12)$ & $0.028110(49)$ & $5.018(10)\times 10^{-7}$ & $8.170(18)\times 10^{-11}$ \\
13 & $-0.04854(12)$ & $-0.027576(38)$ & $4.29(17)\times 10^{-8}$ & $1.544(30)\times 10^{-11}$ \\
14 & $0.09522(14)$ & $0.021594(67)$ & $4.941(17)\times 10^{-7}$ & $8.172(30)\times 10^{-11}$ \\
17 & $-0.000793(80)$ & $0.007631(52)$ & $1.33(18)\times 10^{-8}$ & $8.1(30)\times 10^{-13}$ \\
18 & $0.047200(92)$ & $0.009504(25)$ & $1.404(13)\times 10^{-7}$ & $2.289(21)\times 10^{-11}$ \\
$\sum_j \int F_j(z)dz$ & $0.27727(37)$ & $0.08149(15)$ & $1.5969(42)\times 10^{-6}$ & $2.6469(72)\times 10^{-10}$ \\
$\sum_j \int |F_j(z)|dz$ & $0.801$ & $0.427$ & $2.777\times 10^{-5}$ & $8.347\times 10^{-9}$ \\
\end{longtable*}

\begin{longtable*}{ccccc}\caption{Contributions of individual Feynman graphs to $\widetilde{V}_{27}(\vsp{p})$ for several values of $\vsp{p}$ and $(m_0)^2$ from (\ref{eq_operator_vertex}) and (\ref{eq_operator_fse}).  The column ``Object'' refers to the graph number from Fig.~\ref{fig27} or to a special entry. The integral sums do not include graph 1. The momentum $\vsp{p}$ is given in units of $m$; the factor $m^{-2}$ is omitted. (continued)} \\
\hline \hline Object & $|\vsp{p}|=0.001$ & $|\vsp{p}|=1$ & $|\vsp{p}|=1000$ & $|\vsp{p}|=100000$ \\ \hline  \endhead
\caption{Contributions of individual Feynman graphs to $\widetilde{V}_{27}(\vsp{p})$ for several values of $\vsp{p}$ and $(m_0)^2$ from (\ref{eq_operator_vertex}) and (\ref{eq_operator_fse}).  The column ``Object'' refers to the graph number from Fig.~\ref{fig27} or to a special entry. The integral sums do not include graph 1. The momentum $\vsp{p}$ is given in units of $m$; the factor $m^{-2}$ is omitted.}\label{table_v27_individual} \\
\hline \hline Object & $|\vsp{p}|=0.001$ & $|\vsp{p}|=1$ & $|\vsp{p}|=1000$ & $|\vsp{p}|=100000$ \\ \hline  \endfirsthead
\hline \endfoot  \hline \hline \endlastfoot
Value &  &  &  &  \\
\hline
\multicolumn{5}{c}{Contributions independent of $(m_0)^2$} \\
\hline 
1 & $1.1794(23)\times 10^{-10}$ & $8.945(18)\times 10^{-5}$ & $5.966(14)\times 10^{-8}$ & $1.0446(24)\times 10^{-11}$ \\
2 & $0.007853(18)$ & $0.006977(16)$ & $3.704(13)\times 10^{-7}$ & $6.387(24)\times 10^{-11}$ \\
3 & $-0.006212(19)$ & $-0.005628(18)$ & $-3.617(14)\times 10^{-7}$ & $-6.407(27)\times 10^{-11}$ \\
4 & $0.015359(21)$ & $0.013703(19)$ & $7.603(14)\times 10^{-7}$ & $1.3133(26)\times 10^{-10}$ \\
5 & $-0.004893(13)$ & $-0.004450(14)$ & $-2.889(10)\times 10^{-7}$ & $-5.114(19)\times 10^{-11}$ \\
6 & $-0.000838(50)$ & $-0.000707(36)$ & $-1.02(22)\times 10^{-8}$ & $-0.29(41)\times 10^{-12}$ \\
7 & $0.000694(45)$ & $0.000671(32)$ & $0.37(23)\times 10^{-8}$ & $0.72(42)\times 10^{-12}$ \\
8 & $-0.000766(34)$ & $-0.000667(24)$ & $-0.71(17)\times 10^{-8}$ & $-1.23(31)\times 10^{-12}$ \\
9 & $-0.000930(57)$ & $-0.000842(40)$ & $-0.39(22)\times 10^{-8}$ & $-1.53(40)\times 10^{-12}$ \\
10 & $0.000379(25)$ & $0.000337(24)$ & $0.38(15)\times 10^{-8}$ & $0.19(29)\times 10^{-12}$ \\
13 & $-0.020173(24)$ & $-0.016906(21)$ & $-3.8441(78)\times 10^{-7}$ & $-6.182(11)\times 10^{-11}$ \\
14 & $0.024843(17)$ & $0.021066(16)$ & $5.5285(53)\times 10^{-7}$ & $8.9900(71)\times 10^{-11}$ \\
15 & $-0.024895(22)$ & $-0.021124(18)$ & $-5.6629(64)\times 10^{-7}$ & $-9.1851(87)\times 10^{-11}$ \\
16 & $0.025353(20)$ & $0.021492(19)$ & $5.5004(53)\times 10^{-7}$ & $8.9184(79)\times 10^{-11}$ \\
17 & $-0.018383(19)$ & $-0.015482(20)$ & $-3.7278(68)\times 10^{-7}$ & $-5.9899(88)\times 10^{-11}$ \\
21 & $-0.002905(20)$ & $-0.002711(18)$ & $-1.3917(63)\times 10^{-7}$ & $-2.2685(85)\times 10^{-11}$ \\
22 & $0.006321(13)$ & $0.005492(11)$ & $2.0150(45)\times 10^{-7}$ & $3.2778(62)\times 10^{-11}$ \\
23 & $-0.017323(19)$ & $-0.014870(18)$ & $-4.5507(56)\times 10^{-7}$ & $-7.4009(80)\times 10^{-11}$ \\
24 & $0.008040(10)$ & $0.006883(13)$ & $1.9615(33)\times 10^{-7}$ & $3.1804(56)\times 10^{-11}$ \\
27 & $-0.009463(19)$ & $-0.007995(17)$ & $-2.2972(67)\times 10^{-7}$ & $-3.7245(82)\times 10^{-11}$ \\
28 & $0.0098611(95)$ & $0.008464(12)$ & $2.6157(38)\times 10^{-7}$ & $4.2864(48)\times 10^{-11}$ \\
$\sum_j \int |F_j(z)|dz$ & $0.584$ & $0.513$ & $4.95\times 10^{-5}$ & $1.29\times 10^{-8}$ \\
\hline
\multicolumn{5}{c}{Dependent on $(m_0)^2$ contributions, $(m_0)^2=m^2$} \\
\hline 
11 & $0.00748(15)$ & $0.00593(12)$ & $2.40(27)\times 10^{-8}$ & $-0.15(45)\times 10^{-12}$ \\
12 & $0.004128(92)$ & $0.003125(94)$ & $-2.42(28)\times 10^{-8}$ & $-1.20(49)\times 10^{-12}$ \\
18 & $0.006680(57)$ & $0.005296(58)$ & $1.015(15)\times 10^{-7}$ & $1.767(29)\times 10^{-11}$ \\
19 & $-0.006867(95)$ & $-0.00522(11)$ & $1.362(26)\times 10^{-7}$ & $2.495(45)\times 10^{-11}$ \\
20 & $-0.00423(11)$ & $-0.00335(10)$ & $1.30(27)\times 10^{-8}$ & $5.82(48)\times 10^{-12}$ \\
25 & $0.01538(12)$ & $0.01254(10)$ & $7.15(26)\times 10^{-8}$ & $6.91(46)\times 10^{-12}$ \\
26 & $-0.00550(13)$ & $-0.004815(81)$ & $-1.389(26)\times 10^{-7}$ & $-2.079(49)\times 10^{-11}$ \\
29 & $0.00058(11)$ & $0.00083(12)$ & $1.240(26)\times 10^{-7}$ & $1.993(46)\times 10^{-11}$ \\
30 & $-0.00125(10) $ & $-0.000928(51)$ & $0.85(18)\times 10^{-8}$ & $3.09(33)\times 10^{-12}$ \\
$\sum_j \int F_j(z)dz$ & $0.01639(33)$ & $0.01341(28)$ & $3.156(74)\times 10^{-7}$ & $5.62(13)\times 10^{-11}$ \\
$\sum_j \int |F_j(z)|dz$ & $0.711$ & $0.627$ & $5.553\times 10^{-5}$ & $1.362\times 10^{-8}$ \\
\hline
\multicolumn{5}{c}{Dependent on $(m_0)^2$ contributions, $(m_0)^2=0$} \\
\hline 
11 & $0.006066(80)$ & $0.004853(80)$ & $1.52(23)\times 10^{-8}$ & $0.20(38)\times 10^{-12}$ \\
12 & $0.005446(67)$ & $0.004098(82)$ & $-0.77(25)\times 10^{-8}$ & $-0.33(42)\times 10^{-12}$ \\
18 & $0.007296(28)$ & $0.005918(31)$ & $1.083(14)\times 10^{-7}$ & $1.806(25)\times 10^{-11}$ \\
19 & $-0.008153(93)$ & $-0.006309(68)$ & $1.254(23)\times 10^{-7}$ & $2.342(39)\times 10^{-11}$ \\
20 & $-0.002987(92)$ & $-0.002352(74)$ & $2.70(24)\times 10^{-8}$ & $5.94(42)\times 10^{-12}$ \\
25 & $0.01389(11)$ & $0.011378(47)$ & $5.69(23)\times 10^{-8}$ & $5.63(39)\times 10^{-12}$ \\
26 & $-0.00402(12)$ & $-0.003830(56)$ & $-1.275(24)\times 10^{-7}$ & $-2.061(43)\times 10^{-11}$ \\
29 & $-0.00077(10)$ & $-0.000276(78)$ & $1.090(23)\times 10^{-7}$ & $1.904(39)\times 10^{-11}$ \\
30 & $-0.000482(46)$ & $-0.000364(37)$ & $1.42(17)\times 10^{-8}$ & $2.98(29)\times 10^{-12}$ \\
$\sum_j \int F_j(z)dz$ & $0.01628(26)$ & $0.01312(19)$ & $3.209(66)\times 10^{-7}$ & $5.43(11)\times 10^{-11}$ \\
$\sum_j \int |F_j(z)|dz$ & $0.513$ & $0.454$ & $4.78\times 10^{-5}$ & $1.229\times 10^{-8}$ \\
\hline
\multicolumn{5}{c}{Dependent on $(m_0)^2$ contributions, $(m_0)^2=-m^2$} \\
\hline 
11 & $0.005565(92)$ & $0.004565(77)$ & $0.92(22)\times 10^{-8}$ & $-1.05(37)\times 10^{-12}$ \\
12 & $0.005898(94)$ & $0.004393(62)$ & $-0.48(25)\times 10^{-8}$ & $-0.88(40)\times 10^{-12}$ \\
18 & $0.007509(35)$ & $0.006079(31)$ & $1.104(13)\times 10^{-7}$ & $1.845(24)\times 10^{-11}$ \\
19 & $-0.008676(69)$ & $-0.006622(56)$ & $1.165(22)\times 10^{-7}$ & $2.248(37)\times 10^{-11}$ \\
20 & $-0.002405(58)$ & $-0.001772(60)$ & $3.31(23)\times 10^{-8}$ & $6.40(40)\times 10^{-12}$ \\
25 & $0.013572(53)$ & $0.010987(71)$ & $5.32(22)\times 10^{-8}$ & $5.30(37)\times 10^{-12}$ \\
26 & $-0.003766(76)$ & $-0.003331(90)$ & $-1.267(24)\times 10^{-7}$ & $-1.926(41)\times 10^{-11}$ \\
29 & $-0.001186(92)$ & $-0.000783(62)$ & $1.083(22)\times 10^{-7}$ & $1.784(37)\times 10^{-11}$ \\
30 & $-0.000245(43)$ & $-0.000150(35)$ & $1.50(16)\times 10^{-8}$ & $3.16(27)\times 10^{-12}$ \\
$\sum_j \int F_j(z)dz$ & $0.01627(21)$ & $0.01337(19)$ & $3.143(64)\times 10^{-7}$ & $5.24(11)\times 10^{-11}$ \\
$\sum_j \int |F_j(z)|dz$ & $0.453$ & $0.4$ & $4.526\times 10^{-5}$ & $1.185\times 10^{-8}$ \\
\hline
\multicolumn{5}{c}{Dependent on $(m_0)^2$ contributions, $(m_0)^2=-5m^2$} \\
\hline 
11 & $0.004862(28)$ & $0.003847(25)$ & $0.26(14)\times 10^{-8}$ & $-1.74(25)\times 10^{-12}$ \\
12 & $0.006725(26)$ & $0.005180(26)$ & $0.44(16)\times 10^{-8}$ & $0.38(28)\times 10^{-12}$ \\
18 & $0.007949(21)$ & $0.006428(17)$ & $1.1396(92)\times 10^{-7}$ & $1.869(17)\times 10^{-11}$ \\
19 & $-0.009455(27)$ & $-0.007277(23)$ & $1.144(15)\times 10^{-7}$ & $2.202(26)\times 10^{-11}$ \\
20 & $-0.001674(27)$ & $-0.001265(26)$ & $3.55(16)\times 10^{-8}$ & $7.52(29)\times 10^{-12}$ \\
25 & $0.012717(28)$ & $0.010283(22)$ & $4.86(15)\times 10^{-8}$ & $4.62(26)\times 10^{-12}$ \\
26 & $-0.002910(29)$ & $-0.002698(29)$ & $-1.183(17)\times 10^{-7}$ & $-1.862(29)\times 10^{-11}$ \\
29 & $-0.002036(29)$ & $-0.001378(25)$ & $9.98(15)\times 10^{-8}$ & $1.738(26)\times 10^{-11}$ \\
30 & $0.000136(18)$ & $0.000165(19)$ & $1.80(12)\times 10^{-8}$ & $3.48(20)\times 10^{-12}$ \\
$\sum_j \int F_j(z)dz$ & $0.016314(78)$ & $0.013285(71)$ & $3.189(43)\times 10^{-7}$ & $5.371(76)\times 10^{-11}$ \\
$\sum_j \int |F_j(z)|dz$ & $0.361$ & $0.316$ & $4.05\times 10^{-5}$ & $1.1\times 10^{-8}$ \\
\hline
\multicolumn{5}{c}{Dependent on $(m_0)^2$ contributions, $(m_0)^2=-50m^2$} \\
\hline 
11 & $0.003032(76)$ & $0.002382(43)$ & $-1.05(15)\times 10^{-8}$ & $-2.70(27)\times 10^{-12}$ \\
12 & $0.008491(54)$ & $0.006723(76)$ & $1.98(18)\times 10^{-8}$ & $1.27(31)\times 10^{-12}$ \\
18 & $0.008866(69)$ & $0.007197(38)$ & $1.2086(92)\times 10^{-7}$ & $1.945(18)\times 10^{-11}$ \\
19 & $-0.011281(71)$ & $-0.008768(33)$ & $1.004(16)\times 10^{-7}$ & $2.066(27)\times 10^{-11}$ \\
20 & $0.000086(80)$ & $0.000307(42)$ & $5.09(18)\times 10^{-8}$ & $8.83(31)\times 10^{-12}$ \\
25 & $0.010991(50)$ & $0.008790(48)$ & $3.11(16)\times 10^{-8}$ & $3.24(28)\times 10^{-12}$ \\
26 & $-0.001122(55)$ & $-0.001157(78)$ & $-1.036(18)\times 10^{-7}$ & $-1.709(31)\times 10^{-11}$ \\
29 & $-0.003837(50)$ & $-0.002972(52)$ & $8.41(16)\times 10^{-8}$ & $1.643(28)\times 10^{-11}$ \\
30 & $0.001063(23)$ & $0.000914(30)$ & $2.50(13)\times 10^{-8}$ & $4.41(22)\times 10^{-12}$ \\
$\sum_j \int F_j(z)dz$ & $0.01629(18)$ & $0.01341(15)$ & $3.181(47)\times 10^{-7}$ & $5.452(82)\times 10^{-11}$ \\
$\sum_j \int |F_j(z)|dz$ & $0.353$ & $0.297$ & $3.03\times 10^{-5}$ & $9.24\times 10^{-9}$ 
\end{longtable*}

\section{Conclusion}

Momentum-space values for the expansion of the vacuum polarization potential in the Coulomb field of a pointlike source up to order $\alpha^2 (Z\alpha)^7$ were presented in ~\cite{vp_coulomb_potential_2025} in tabular form. Their application to high-precison calculations of energy levels of atomic systems was also discussed there. These values agree well with the result of order $\alpha^2 (Z\alpha)^3$ at $\vsp{p}=0$ from ~\cite{lee_krachkov_2023}. The values of orders $\alpha^2 (Z\alpha)^5$ and $\alpha^2 (Z\alpha)^7$, as well as those of order $\alpha^2 (Z\alpha)^3$ at $\vsp{p}\neq 0$ were presented for the first time. The methodology of the calculation is explained in this work. 

The method relies on a reduction to free QED Feynman graphs (with up to 8 loops), the author's approach to removing divergences and reducing to finite integrals, and Monte Carlo integration.

The reduction to free QED (unfolding) is effective for evaluating the vacuum polarization potential in a Coulomb field, but its extension to general bound-state problems is not feasible. The reduction to finite integrals is based on an almost direct application of BPHZ renormalization in Feynman parametric space. A slight modification of BPHZ addresses specific features of free QED Feynman graphs arising in bound-state QED and incorporates ideas from the author's previous work on the free electron $g-2$ calculation.

A nonadaptive Monte Carlo algorithm was employed. It constructs probability density functions from the combinatorial structure of the Feynman graphs. The algorithm has 10 parameters that were preadjusted prior to integration. A wide range of external momenta was taken into account. GPUs were used to carry out the integrations. The  integration algorithm itself, along with its computer implementation, was adapted from the author's earlier work on the free-electron $g-2$ calculation. Technical details are provided.

The contributions of individual Feynman graphs are reported together with Monte Carlo integration statistics. This information can be used for verification. In addition, it shows how efficient the method is and how large the cancellations are (between integrals and within an integral) across the full range of external momenta and also how it depends on the intermediate renormalization.

\begin{acknowledgments}
The author thanks V.~A.~Yerokhin for fruitful recommendations concerning the method and text and also for the problem formulation. Also, the author is grateful to Z.~Harman and C.~H.~Keitel for collaboration and important assistance. All computations were accomplished on the  computing cluster of the Max Planck Institute for Nuclear Physics (Heidelberg, Germany). All pictures were drawn using \texttt{Jaxodraw}.
\end{acknowledgments}

\ \\ 

\bibliography{all_phys_cd2025}

\end{document}